\documentclass[journal,onecolumn,draftcls,12pt]{IEEEtran}

\newlength{\figwidth}
\usepackage{xspace}
\usepackage{url}
\usepackage[cmex10]{amsmath}
\usepackage{bbm}
\usepackage{graphicx}
\usepackage{epstopdf}
\usepackage{paralist}
\usepackage[normalem]{ulem}
\usepackage{fancyref} 
\usepackage{amsmath}
\usepackage{amssymb}
\usepackage[stretch=16,shrink=16,step=4]{microtype}
\usepackage{xcolor}
\definecolor{links}{rgb}{0.7,0,0}   
\definecolor{urls}{rgb}{0,0,0.8}    
\definecolor{cites}{rgb}{0,0,0.8}   
\usepackage{dsfont}
\usepackage{footnote}
\usepackage{balance}
\usepackage{float}
\usepackage{cite}
\usepackage{wasysym}
\usepackage{amssymb}
\let\labelindent\relax
\usepackage[inline]{enumitem}
\usepackage{setspace}
\usepackage{tabularx,booktabs}
\newcolumntype{L}[1]{>{\raggedright\let\newline\\\arraybackslash\hspace{0pt}}m{#1}}
\newcolumntype{C}[1]{>{\centering\let\newline\\\arraybackslash\hspace{0pt}}m{#1}}
\newcolumntype{R}[1]{>{\raggedleft\let\newline\\\arraybackslash\hspace{0pt}}m{#1}}
\usepackage{boldline}

\usepackage{algorithm,algpseudocode}

\usepackage{vmr-symbols-vecbold}
\usepackage{standard-macros}
\usepackage{mathbbol}

\usepackage{tikz}
\usetikzlibrary{patterns}
\usetikzlibrary{shapes,arrows,positioning}
\usetikzlibrary{calc}
\usetikzlibrary{fit}
\usetikzlibrary{plotmarks}
\tikzset{every picture/.style={font issue=\scriptsize, >=stealth},font issue/.style={execute at begin picture={#1\selectfont}}}
\tikzset{three sided left/.style={
        draw=none,
        xshift=\pgflinewidth,
        append after command={
            [shorten <= -0.5\pgflinewidth]
            ([shift={(-1.5\pgflinewidth,-0.5\pgflinewidth)}]\tikzlastnode.north east) edge ([shift={( 0.5\pgflinewidth,-0.5\pgflinewidth)}]\tikzlastnode.north west) 
            ([shift={( 0.5\pgflinewidth,-0.5\pgflinewidth)}]\tikzlastnode.north west) edge ([shift={( 0.5\pgflinewidth,+0.5\pgflinewidth)}]\tikzlastnode.south west)            
            ([shift={( 0.5\pgflinewidth,+0.5\pgflinewidth)}]\tikzlastnode.south west) edge ([shift={(-1.0\pgflinewidth,+0.5\pgflinewidth)}]\tikzlastnode.south east)
        }}}
        
\tikzset{three sided right/.style={
        draw=none,
        xshift=-\pgflinewidth,
        append after command={
            [shorten <= -0.5\pgflinewidth]
            ([shift={( 1.5\pgflinewidth,-0.5\pgflinewidth)}]\tikzlastnode.north west) edge ([shift={(-0.5\pgflinewidth,-0.5\pgflinewidth)}]\tikzlastnode.north east) 
            ([shift={(-0.5\pgflinewidth,-0.5\pgflinewidth)}]\tikzlastnode.north east) edge ([shift={(-0.5\pgflinewidth,+0.5\pgflinewidth)}]\tikzlastnode.south east)            
            ([shift={(-0.5\pgflinewidth,+0.5\pgflinewidth)}]\tikzlastnode.south east) edge ([shift={( 1.0\pgflinewidth,+0.5\pgflinewidth)}]\tikzlastnode.south west)
        }}}

\usepackage{pgfplots}
\usepgfplotslibrary{fillbetween}
\pgfplotsset{
  compat=newest, 
  width=\columnwidth,    
  height=0.8\columnwidth,   
  plot coordinates/math parser=false,
  standard/.style={
    axis equal,
    axis line style=help lines,
    axis x line=center,
    axis y line=center,
    axis z line=center},
    grid style={dashed,gray},
    minor grid style={dotted,gray},
    major grid style={dotted,gray},
    ylabel absolute, ylabel style={yshift=-0.4cm},
    xlabel absolute, xlabel style={yshift=0.25cm}
}

\makeatletter
 \pgfdeclarepatternformonly[\tikz@pattern@color,\LineSpace,\LineWidth]{my horizontal lines}%
    {\pgfpointorigin}{\pgfqpoint{100pt}{1pt}}{\pgfqpoint{100pt}{\LineSpace}}%
    {
        \pgfsetcolor{\tikz@pattern@color}
        \pgfsetlinewidth{\LineWidth}
        \pgfpathmoveto{\pgfqpoint{0pt}{0.5pt}}
        \pgfpathlineto{\pgfqpoint{100pt}{0.5pt}}
        \pgfusepath{stroke}
    }
 
 \pgfdeclarepatternformonly[\tikz@pattern@color,\LineSpace,\LineWidth]{my vertical lines}%
    {\pgfpointorigin}{\pgfqpoint{1pt}{100pt}}{\pgfqpoint{\LineSpace}{100pt}}%
    {
        \pgfsetcolor{\tikz@pattern@color}
        \pgfsetlinewidth{\LineWidth}
        \pgfpathmoveto{\pgfqpoint{0.5pt}{0pt}}
        \pgfpathlineto{\pgfqpoint{0.5pt}{100pt}}
        \pgfusepath{stroke}
    } 
 
 \pgfdeclarepatternformonly[\tikz@pattern@color,\LineSpace,\LineWidth]{my grid}%
    {\pgfqpoint{-1pt}{-1pt}}{\pgfqpoint{\LineSpace}{\LineSpace}}
    {\pgfqpoint{\LineSpace}{\LineSpace}}%
    {
        \pgfsetcolor{\tikz@pattern@color}
        \pgfsetlinewidth{\LineWidth}
        \pgfpathmoveto{\pgfqpoint{0pt}{0pt}}
        \pgfpathlineto{\pgfqpoint{0pt}{\LineSpace + 0.1pt}}
        \pgfpathmoveto{\pgfqpoint{0pt}{0pt}}
        \pgfpathlineto{\pgfqpoint{\LineSpace + 0.1pt}{0pt}}
        \pgfusepath{stroke}
    }
 
 \pgfdeclarepatternformonly[\tikz@pattern@color,\LineSpace,\LineWidth]{my north east lines}
    {\pgfqpoint{-\LineWidth}{-\LineWidth}}{\pgfqpoint{\LineSpace}{\LineSpace}}
    {\pgfqpoint{\LineSpace}{\LineSpace}}%
    {
        \pgfsetcolor{\tikz@pattern@color}
        \pgfsetlinewidth{\LineWidth}
        \pgfpathmoveto{\pgfqpoint{-\LineWidth}{-\LineWidth}}
        \pgfpathlineto{\pgfqpoint{\LineSpace + 0.1pt}{\LineSpace + 0.1pt}}
        \pgfusepath{stroke}
    }
 
 \pgfdeclarepatternformonly[\tikz@pattern@color,\LineSpace,\LineWidth]{my north west lines}
    {\pgfqpoint{-\LineWidth}{-\LineWidth}}{\pgfqpoint{\LineSpace}{\LineSpace}}
    {\pgfqpoint{\LineSpace}{\LineSpace}}%
    {
        \pgfsetcolor{\tikz@pattern@color}
        \pgfsetlinewidth{\LineWidth}
        \pgfpathmoveto{\pgfqpoint{-\LineWidth}{\LineSpace}}
        \pgfpathlineto{\pgfqpoint{\LineSpace + 0.1pt}{-\LineWidth}}
        \pgfusepath{stroke}
    }
 
 \pgfdeclarepatternformonly[\tikz@pattern@color,\LineSpace,\LineWidth]{my crosshatch}%
    {\pgfqpoint{-1pt}{-1pt}}{\pgfqpoint{\LineSpace}{\LineSpace}}
    {\pgfqpoint{\LineSpace}{\LineSpace}}%
    {
        \pgfsetcolor{\tikz@pattern@color}
        \pgfsetlinewidth{\LineWidth}
        \pgfpathmoveto{\pgfqpoint{\LineSpace + 0.1pt}{0pt}}
        \pgfpathlineto{\pgfqpoint{0pt}{\LineSpace + 0.1pt}}
        \pgfpathmoveto{\pgfqpoint{0pt}{0pt}}
        \pgfpathlineto{\pgfqpoint{\LineSpace + 0.1pt}{\LineSpace + 0.1pt}}
        \pgfusepath{stroke}
    }
 
 \pgfdeclarepatternformonly[\tikz@pattern@color,\LineSpace,\PointSize]{my dots}%
    {\pgfqpoint{-\LineSpace*0.25}{-\LineSpace*0.25}}
    {\pgfqpoint{\LineSpace*0.25}{\LineSpace*0.25}}
    {\pgfqpoint{\LineSpace*0.75}{\LineSpace*0.75}}%
    {
        \pgfsetcolor{\tikz@pattern@color}
        \pgfpathcircle{\pgfqpoint{0pt}{0pt}}{\PointSize}
        \pgfusepath{fill}
    }
\makeatother
 
\newdimen\LineSpace
\newdimen\PointSize
\newdimen\LineWidth
\tikzset{
    line space/.code={\LineSpace=#1},
    line space=3pt
}
\tikzset{
    point size/.code={\PointSize=#1},
    point size=.5pt
}
\tikzset{
    pattern line width/.code={\LineWidth=#1},
    pattern line width=.4pt
}

\DeclareSymbolFontAlphabet{\amsmathbb}{AMSb}%

\newcommand{\lro}[1]{\lefto({#1}\right)}																
\newcommand{\lrbo}[1]{\lefto \lbrace {#1} \right \rbrace}															
\newcommand{\lrho}[1]{\lefto [ {#1} \right ]}																				

\newcommand{\lr}[1]{\left({#1}\right)}																
\newcommand{\lrb}[1]{\left \lbrace {#1} \right \rbrace}															

\safemath{\dopplerspread}{B_D}																								
\safemath{\delayspread}{T_D}																									
\safemath{\nc}{n\sub{c}}																										
\safemath{\nf}{n\sub{f}}																										
\safemath{\efa}{p\sub{sc}}
\safemath{\efb}{p\sub{cs}}
\safemath{\ef}{\epsilon\sub{f}	}
\safemath{\nd}{n\sub{d}}																										
\safemath{\ntx}{n\sub{t}} 																											
\safemath{\nrx}{n\sub{r}}																											
\safemath{\ntxt}{\tilde{n\sub{t}}}																											
\safemath{\cb}{\ensuremath{L}} 																								
\safemath{\cl}{\ensuremath{n}} 																								
\safemath{\txanto}{{\ensuremath{\tilde{m}_t}}} 																		
\safemath{\cs}{M} 																														
\safemath{\idPustm}{\ensuremath{S_{k}}}
\safemath{\error}{\ensuremath{\epsilon}} 																				
\safemath{\eexp}{\ensuremath{\mathcal{E}}} 																			
\safemath{\nsubc}{n\sub{s}}			 																						
\safemath{\nofdm}{n\sub{o}} 																									
\safemath{\bc}{\ensuremath{B_c}} 																							
\safemath{\ts}{\ensuremath{T_s}} 																							
\safemath{\nrb}{\ensuremath{n_{rb}}} 																						
\safemath{\nres}{\ell}
\safemath{\nr}{n\sub{r}}
\newcommand{\cgauss}[2]{\mathcal{CN}\lro{\ensuremath{#1, #2}  }}   								
\safemath{\maxk}{M^*\lr{\nres, \nsubc, \nofdm, \epsilon, \rho}}
\safemath{\Rmax}{R^*}
\safemath{\Emin}{E\sub{b}^*/N_0}
\safemath{\Eminf}{\frac{E\sub{b}^*}{N_0}}
\safemath{\np}{\ensuremath{n\sub{p}}}
\safemath{\ndf}{\ensuremath{\bar{n}\sub{d}}}
\safemath{\npf}{\ensuremath{\bar{n}\sub{p}}}
\safemath{\code}{\ensuremath{\mathcal{C}}}
\safemath{\err}{\ensuremath{\epsilon}}
\safemath{\rp}{\ensuremath{\rho\sub{p}}}
\safemath{\rd}{\ensuremath{\rho\sub{d}}}
\safemath{\cohtime}{\ensuremath{T\sub{c}}}
\safemath{\cohbw}{\ensuremath{B\sub{c}}}
\safemath{\nmax}{\ensuremath{\ell\sub{m}}}
\safemath{\ntot}{\ensuremath{n\sub{tot}}}

\safemath{\yp}{\ensuremath{\randvecy_{\nu}^{(\text{p})}}}
\safemath{\yd}{\ensuremath{\randvecy_{\nu}^{(\text{d})}}}
\safemath{\ypd}{\ensuremath{\vecy_{\nu}^{(\text{p})}}}
\safemath{\ydd}{\ensuremath{\vecy_{\nu}^{(\text{d})}}}

\safemath{\ypf}{\ensuremath{\bar{\randvecy}_{\nu}^{(\text{p})}}}
\safemath{\ydf}{\ensuremath{\bar{\randvecy}_{\nu}^{(\text{d})}}}
\safemath{\ypdf}{\ensuremath{\bar{\vecy}_{\nu}^{(\text{p})}}}
\safemath{\yddf}{\ensuremath{\bar{\vecy}_{\nu}^{(\text{d})}}}

\safemath{\xp}{\ensuremath{\vecx^{(\text{p})}}}
\safemath{\xd}{\ensuremath{\randvecx_{\nu}^{(\text{d})}}}
\safemath{\xdd}{\ensuremath{\vecx_{\nu}^{(\text{d})}}}

\safemath{\xpf}{\ensuremath{\bar{\vecx}^{(\text{p})}}}
\safemath{\xdf}{\ensuremath{\bar{\randvecx}_{\nu}^{(\text{d})}}}
\safemath{\xddf}{\ensuremath{\bar{\vecx}_{\nu}^{(\text{d})}}}

\safemath{\xdb}{\ensuremath{\overline{\randvecx}^{(\text{d})}}}
\safemath{\Pxd}{\ensuremath{P_{\randvecx^{(\text{d})}}}}

\safemath{\xpbar}{\ensuremath{\overline{\matX}^{(\text{p})}}}
\safemath{\xdbar}{\ensuremath{\overline{\randmatX}^{(\text{d})}}}

\safemath{\xdv}{\ensuremath{\randvecx^{(\text{d})}}}
\safemath{\xdbarv}{\ensuremath{\overline{\randvecx}^{(\text{d})}}}
\safemath{\ydv}{\ensuremath{\randvecy^{(\text{d})}}}

\safemath{\xdr}{\ensuremath{\matX^{(\text{d})}}}

\safemath{\ttx}{\ensuremath{\tau\sub{tx}}}
\safemath{\trx}{\ensuremath{\tau\sub{rx}}}
\safemath{\ack}{\ensuremath{\mathrm{s}}}
\safemath{\nack}{\ensuremath{\mathrm{c}}}

\newcommand{\prob}[1]{\ensuremath{\mathbb{P}\lrho{#1}}}

\safemath{\mI}{\ensuremath{i\lro{\randvecy ; \randvecx}}} 				

\safemath{\randveca}{\bm{A}}
\safemath{\randvecb}{\bm{B}}
\safemath{\randvecc}{\bm{C}}
\safemath{\randvecd}{\bm{D}}
\safemath{\randvece}{\bm{E}}
\safemath{\randvecf}{\bm{F}}
\safemath{\randvecg}{\bm{G}}
\safemath{\randvech}{\bm{H}}
\safemath{\randveci}{\bm{I}}
\safemath{\randvecj}{\bm{J}}
\safemath{\randveck}{\bm{K}}
\safemath{\randvecl}{\bm{L}}
\safemath{\randvecm}{\bm{M}}
\safemath{\randvecn}{\bm{N}}
\safemath{\randveco}{\bm{O}}
\safemath{\randvecp}{\bm{P}}
\safemath{\randvecq}{\bm{Q}}
\safemath{\randvecr}{\bm{R}}
\safemath{\randvecs}{\bm{S}}
\safemath{\randvect}{\bm{T}}
\safemath{\randvecu}{\bm{U}}
\safemath{\randvecv}{\bm{V}}
\safemath{\randvecw}{\bm{W}}
\safemath{\randvecx}{\bm{X}}
\safemath{\randvecy}{\bm{Y}}
\safemath{\randvecz}{\bm{Z}}
\safemath{\randvecphi}{\bm{\Phi}}

\safemath{\randmatA}{\amsmathbb{A}}
\safemath{\randmatB}{\amsmathbb{B}}
\safemath{\randmatC}{\amsmathbb{C}}
\safemath{\randmatD}{\amsmathbb{D}}
\safemath{\randmatE}{\amsmathbb{E}}
\safemath{\randmatF}{\amsmathbb{F}}
\safemath{\randmatG}{\amsmathbb{G}}
\safemath{\randmatH}{\amsmathbb{H}}
\safemath{\randmatI}{\amsmathbb{I}}
\safemath{\randmatJ}{\amsmathbb{J}}
\safemath{\randmatK}{\amsmathbb{K}}
\safemath{\randmatL}{\amsmathbb{L}}
\safemath{\randmatM}{\amsmathbb{M}}
\safemath{\randmatN}{\amsmathbb{N}}
\safemath{\randmatO}{\amsmathbb{O}}
\safemath{\randmatP}{\amsmathbb{P}}
\safemath{\randmatQ}{\amsmathbb{Q}}
\safemath{\randmatR}{\amsmathbb{R}}
\safemath{\randmatS}{\amsmathbb{S}}
\safemath{\randmatT}{\amsmathbb{T}}
\safemath{\randmatU}{\amsmathbb{U}}
\safemath{\randmatV}{\amsmathbb{V}}
\safemath{\randmatW}{\amsmathbb{W}}
\safemath{\randmatX}{\amsmathbb{X}}
\safemath{\randmatY}{\amsmathbb{Y}}
\safemath{\randmatZ}{\amsmathbb{Z}}
\safemath{\randmatSigma}{\mathbb{\Sigma}}
\safemath{\randmatPhi}{\mathbb{\Phi}}
\safemath{\randmatLambda}{\mathbb{\Lambda}}

\safemath{\matSigma}{\bm{\Sigma}}
\safemath{\matPhi}{\bm{\Phi}}
\safemath{\matLambda}{\bm{\Lambda}}

\usepackage{glossaries}
\newglossaryentry{biawgn}
{
  name={bi-AWGN},
  description={binary-input additive white Gaussian noise},
  first={\glsentrydesc{biawgn} (\glsentrytext{biawgn})}
}

\newglossaryentry{flnf}
{
  name={FLNF},
  description={fixed-length no-feedback},
  first={\glsentrydesc{flnf} (\glsentrytext{flnf})}
}

\newglossaryentry{crc}
{
  name={CRC},
  description={cyclic redundancy check},
  first={\glsentrydesc{crc} (\glsentrytext{crc})}
}

\newglossaryentry{bsc}
{
  name={BSC},
  description={binary symmetric channel},
  first={\glsentrydesc{bsc} (\glsentrytext{bsc})}
  plural={BSCs},
  descriptionplural={binary symmetric channels},
  firstplural={\glsentrydescplural{bsc} (\glsentryplural{bsc})}
}

\newglossaryentry{bec}
{
  name={BEC},
  description={binary-erasure channel},
  first={\glsentrydesc{bec} (\glsentrytext{bec})}
}
\newglossaryentry{awgn}
{
  name={AWGN},
  description={additive white Gaussian noise},
  first={\glsentrydesc{awgn} (\glsentrytext{awgn})}
}
\newglossaryentry{3gpp}
{
  name={3GPP},
  description={3rd generation partnership project},
  first={\glsentrydesc{3gpp} (\glsentrytext{3gpp})}
}

\newglossaryentry{dl}
{
  name={DL},
  description={downlink},
  first={\glsentrydesc{dl} (\glsentrytext{dl})}
}

\newglossaryentry{LED}
{
  name={LED},
  description={light emitting diode},
  first={\glsentrydesc{LED} (\glsentrytext{LED})},
  plural={LEDs},
  descriptionplural={light emitting diodes},
  firstplural={\glsentrydescplural{LED} (\glsentryplural{LED})}
}

\newglossaryentry{lte}
{
  name={LTE},
  description={long term evolution},
  first={\glsentrydesc{lte} (\glsentrytext{lte})}
}
\newglossaryentry{ofdm}
{
  name={OFDM},
  description={orthogonal frequency-division multiplexing},
  first={\glsentrydesc{ofdm} (\glsentrytext{ofdm})}
}
\newglossaryentry{ue}
{
  name={UE},
  description={user equipment},
  first={\glsentrydesc{ue} (\glsentrytext{ue})},
  plural={UE's},
  descriptionplural={user equipments},
  firstplural={\glsentrydescplural{ue} (\glsentryplural{ue})}
}
\newglossaryentry{ul}
{
  name={UL},
  description={uplink},
  first={\glsentrydesc{ul} (\glsentrytext{ul})}
}
\newglossaryentry{rb}
{
  name={RB},
  description={resource block},
  first={\glsentrydesc{rb} (\glsentrytext{rb})},
  plural={RBs},
  descriptionplural={resource blocks},
  firstplural={\glsentrydescplural{rb} (\glsentryplural{rb})}
}
\newglossaryentry{cu}
{
  name={CU},
  description={channel use},
  first={\glsentrydesc{cu} (\glsentrytext{cu})},
  plural={CUs},
  descriptionplural={channel uses},
  firstplural={\glsentrydescplural{cu} (\glsentryplural{cu})}
}
\newglossaryentry{tti}
{
  name={TTI},
  description={transmission time interval},
  first={\glsentrydesc{tti} (\glsentrytext{tti})},
  plural={TTI's},
  descriptionplural={transmission time intervals},
  firstplural={\glsentrydescplural{tti} (\glsentryplural{tti})}
}
\newglossaryentry{iid}
{
  name={i.i.d.},
  description={independent and identically distributed},
  first={\glsentrydesc{iid} (\glsentrytext{iid})}
}
\newglossaryentry{psd}
{
  name={PSD},
  description={power spectral density},
  first={\glsentrydesc{psd} (\glsentrytext{psd})}
}

\newglossaryentry{mimo}
{
  name={MIMO},
  description={multiple-input multiple-output},
  first={\glsentrydesc{mimo} (\glsentrytext{mimo})}
}

\newglossaryentry{bler}
{
  name={BLER},
  description={block error probability},
  first={\glsentrydesc{bler} (\glsentrytext{bler})}
}
\newglossaryentry{mtc}
{
  name={MTC},
  description={machine-type communication},
  first={\glsentrydesc{mtc} (\glsentrytext{mtc})}
}
\newglossaryentry{csi}
{
  name={CSI},
  description={channel state information},
  first={\glsentrydesc{csi} (\glsentrytext{csi})}
}
\newglossaryentry{dvbt}
{
  name={DVB-T},
  description={terrestrial digital video broadcasting},
  first={\glsentrydesc{dvbt} (\glsentrytext{dvbt})}
}
\newglossaryentry{pat}
{
  name={PAT},
  description={pilot-assisted transmission},
  first={\glsentrydesc{pat} (\glsentrytext{pat})}
}
\newglossaryentry{ml}
{
  name={ML},
  description={maximum likelihood},
  first={\glsentrydesc{ml} (\glsentrytext{ml})}
}
\newglossaryentry{dt}
{
  name={DT},
  description={dependence-testing},
  first={\glsentrydesc{dt} (\glsentrytext{dt})}
}
\newglossaryentry{ustm}
{
  name={USTM},
  description={unitary space-time modulation},
  first={\glsentrydesc{ustm} (\glsentrytext{ustm})}
}
\newglossaryentry{siso}
{
  name={SISO},
  description={single-input single-output},
  first={\glsentrydesc{siso} (\glsentrytext{siso})}
}
\newglossaryentry{snn}
{
  name={SNN},
  description={scaled nearest-neighbor},
  first={\glsentrydesc{snn} (\glsentrytext{snn})}
}
\newglossaryentry{rcus}
{
  name={RCUs},
  description={random-coding union bound with parameter $s$},
  first={\glsentrydesc{rcus} (\glsentrytext{rcus})}
}
\newglossaryentry{osd}
{
  name={OSD},
  description={ordered statistics decoding},
  first={\glsentrydesc{osd} (\glsentrytext{osd})}
}
\newglossaryentry{llr}
{
  name={LLR},
  description={log-likelihood ratio},
  first={\glsentrydesc{llr} (\glsentrytext{llr})}
   plural={LLR's},
  descriptionplural={log-likelihood ratios},
  firstplural={\glsentrydescplural{llr} (\glsentryplural{llr})}
}
\newglossaryentry{qpsk}
{
  name={QPSK},
  description={quaternary phase shift keying},
  first={\glsentrydesc{qpsk} (\glsentrytext{qpsk})}
}
\newglossaryentry{nlos}
{
  name={NLOS},
  description={non-line-of-sight},
  first={\glsentrydesc{nlos} (\glsentrytext{nlos})}
}
\newglossaryentry{los}
{
  name={LOS},
  description={line-of-sight},
  first={\glsentrydesc{los} (\glsentrytext{los})}
}
\newglossaryentry{cgf}
{
  name={CGF},
  description={cumulant-generating function},
  first={\glsentrydesc{cgf} (\glsentrytext{cgf})}
   plural={CGF's},
  descriptionplural={cumulant-generating functions},
  firstplural={\glsentrydescplural{cgf} (\glsentryplural{cgf})}
}
\newglossaryentry{pdf}
{
  name={PDF},
  description={probability density function},
  first={\glsentrydesc{pdf} (\glsentrytext{pdf})}
   plural={PDF's},
  descriptionplural={probability density functions},
  firstplural={\glsentrydescplural{pdf} (\glsentryplural{pdf})}
}

\newglossaryentry{ccdf}
{
  name={CCDF},
  description={complementary cumulative distribution function},
  first={\glsentrydesc{ccdf} (\glsentrytext{ccdf})}
}

\newglossaryentry{cdf}
{
  name={CDF},
  description={cumulative distribution function},
  first={\glsentrydesc{cdf} (\glsentrytext{cdf})}
}

\newglossaryentry{gmi}
{
  name={GMI},
  description={generalized mutual information},
  first={\glsentrydesc{gmi} (\glsentrytext{gmi})}
}

\newglossaryentry{arq}
{
  name={ARQ},
  description={automatic repeat request},
  first={\glsentrydesc{arq} (\glsentrytext{arq})}
}

\newglossaryentry{harq}
{
  name={HARQ},
  description={hybrid automatic repeat request},
  first={\glsentrydesc{harq} (\glsentrytext{harq})}
}

\newglossaryentry{fbl}
{
  name={FBL-NF},
  description={fixed blocklength with no feedback},
  first={\glsentrydesc{fbl} (\glsentrytext{fbl})}
}
\newglossaryentry{ssnf}
{
  name={SS-NF},
  description={single-shot no-feedback},
  first={\glsentrydesc{ssnf} (\glsentrytext{ssnf})}
}

\newglossaryentry{urllc}
{
  name={URLLC},
  description={ultra-reliable low-latency communications},
  first={\glsentrydesc{urllc} (\glsentrytext{urllc})}
}

\newglossaryentry{vlsf}
{
  name={VLSF},
  description={Variable-length stop-feedback},
  first={\glsentrydesc{vlsf} (\glsentrytext{vlsf})}
}

\newglossaryentry{vlnsf}
{
  name={VLNSF},
  description={variable-length noisy stop feedback},
  first={\glsentrydesc{vlnsf} (\glsentrytext{vlnsf})}
}

\newglossaryentry{dmt}
{
  name={DMT},
  description={diversity-multiplexing tradeoff},
  first={\glsentrydesc{dmt} (\glsentrytext{dmt})}
}

\newglossaryentry{dmdt}
{
  name={DMDT},
  description={diversity-multiplexing-delay tradeoff},
  first={\glsentrydesc{dmdt} (\glsentrytext{dmdt})}
}

\newglossaryentry{tddt}
{
  name={TDDT},
  description={throughput-diversity-delay tradeoff},
  first={\glsentrydesc{tddt} (\glsentrytext{tddt})}
}

\newglossaryentry{stbc}
{
  name={STBC},
  description={space-time block-codes},
  first={\glsentrydesc{stbc} (\glsentrytext{stbc})},
  plural={STBC's},
  descriptionplural={space-time block-codes},
  firstplural={\glsentrydescplural{stbc} (\glsentryplural{stbc})}
}

\newglossaryentry{ostbc}
{
  name={OSTBC},
  description={orthogonal space-time block code},
  first={\glsentrydesc{stbc} (\glsentrytext{stbc})},
  plural={OSTBCs},
  descriptionplural={space-time block-codes},
  firstplural={\glsentrydescplural{ostbc} (\glsentryplural{ostbc})}
}

\newglossaryentry{harqir}
{
  name={HARQ-IR},
  description={hybrid automatic repeat request with incremental redundancy},
  first={\glsentrydesc{harqir} (\glsentrytext{harqir})}
}

\newglossaryentry{harqcc}
{
  name={HARQ-CC},
  description={hybrid automatic repeat request with chase combining},
  first={\glsentrydesc{harqcc} (\glsentrytext{harqcc})}
}

\newglossaryentry{wp1}
{
  name={w.p.1.},
  description={with probability one},
  first={\glsentrydesc{wp1} (\glsentrytext{wp1})}
}

\newglossaryentry{vlff}
{
  name={VLFF},
  description={variable-length full feedback},
  first={\glsentrydesc{vlff} (\glsentrytext{vlff})}
}

\newglossaryentry{dmc}
{
  name={DMC},
  description={discrete memoryless channel},
  first={\glsentrydesc{dmc} (\glsentrytext{dmc})}
  plural={DMCs},
  descriptionplural={discrete memoryless channels},
  firstplural={\glsentrydescplural{dmc} (\glsentryplural{dmc})}
}

\algnewcommand{\Initialize}[1]{%
  \State \textbf{Initialize:}
  \Statex \hspace*{\algorithmicindent}\parbox[t]{.8\linewidth}{\raggedright #1}
}

\usepackage{pgfplots}
\usepgfplotslibrary{groupplots}
\usetikzlibrary{pgfplots.groupplots}

\usepgfplotslibrary{external}
\newcommand\linew{1pt} 
\newcommand{\fwidth}{\textwidth}

\makeatletter
\def\@IEEEinterspaceratioM{0.265}
\def\@IEEEinterspaceMINratioM{0.1651}
\def\@IEEEinterspaceMAXratioM{0.38}

\def\@IEEEinterspaceratioB{0.31}
\def\@IEEEinterspaceMINratioB{0.19}
\def\@IEEEinterspaceMAXratioB{0.38}
\@IEEEtunefonts
\makeatother
\let\abs\undefined
\newcommand{\abs}[1]{\lvert#1\rvert}		

\usepackage[font=scriptsize]{subcaption}
\let\marginnote\undefined

\newcommand{\marginnote}[1]{\marginpar{\framebox{\framebox{#1}}}}

\newcommand{\ju}[1]{{{#1}}} 

\renewcommand{\markblue}[1]{\color{black} #1 }

\allowdisplaybreaks

\begin{document}
\IEEEoverridecommandlockouts
\algnewcommand\algorithmicswitch{\textbf{switch}}
\algnewcommand\algorithmicendswitch{\textbf{end switch}}
\algnewcommand\algorithmiccase{\textbf{case}}
\algdef{SE}[SWITCH]{Switch}{EndSwitch}[1]{\algorithmicswitch\ #1\ \algorithmicdo}{\algorithmicend\ \algorithmicswitch }%
\algdef{SE}[CASE]{Case}{EndCase}[1]{\algorithmiccase\ #1}{\algorithmicend\ \algorithmiccase}%
\algtext*{EndSwitch}%
\algtext*{EndCase}%

\title{Short-packet Transmission via Variable-Length Codes in the Presence of Noisy Stop Feedback}
\author{Johan \"Ostman,~\IEEEmembership{Student Member,~IEEE}, Rahul Devassy, Giuseppe~Durisi,~\IEEEmembership{Senior Member,~IEEE}, Erik~G.~Str\"om,~\IEEEmembership{Senior Member,~IEEE}
\thanks{This work was partly supported by the Swedish Research Council under grants 2014-6066 and 2016-03293.}
\thanks{Parts of the material of this paper have been presented at the IEEE Information Theory Workshop, August 2019, Visby, Sweden~\cite{ostman19-08a}.}
\thanks{Johan \"Ostman, Giuseppe Durisi, and Erik G. Str{\"o}m are with the Department of Electrical Engineering, Chalmers University of Technology, Gothenburg 41296, Sweden (e-mail: \{johanos,durisi,erik.strom\}@chalmers.se).
}
\thanks{Rahul Devassy was with the Department of Electrical Engineering, Chalmers University of Technology, Gothenburg 41296, Sweden (e-mail: devassyhere@gmail.com). }
}
 \maketitle

 \begin{abstract}
   We present an upper bound on the error probability achievable using variable-length stop feedback codes, for a fixed size of the information payload and a given constraint on the maximum latency and the average service time.
   Differently from the bound proposed in Polyanskiy \emph{et al.} (2011), which pertains to the scenario in which the stop signal is sent over a noiseless feedback channel, our bound applies to the practically relevant setup in which the feedback link is noisy.
   \ju{By numerically evaluating our bound, we illustrate that, for fixed latency and reliability constraints, noise in the feedback link can cause a significant increase in the minimum average service time, to the extent that fixed-length codes without feedback may be preferable in some scenarios.}

 \end{abstract}

\section{Introduction}

\gls{vlsf} coding schemes, i.e., schemes such as simple \gls{arq} and \gls{harq}, in which information is transmitted until the reception of a positive acknowledgment (ACK), are ubiquitous in modern wireless communication systems.
This is because they offer a simple yet effective way to adapt the transmission rate to the channel conditions and, hence, reduce the error probability.
The question investigated in this paper is whether such schemes are suitable for \gls{urllc}---one of the new use cases in next-generation wireless systems (5G).

From a physical layer perspective, the URLLC design problem involves answering the following question: can a given information payload be transmitted within a target latency requirement at a desired reliability level?
Unfortunately, classical approaches to answering this question, which rely on large-blocklength results in information theory, are unsuitable whenever the latency requirement is stringent, such as in \gls{urllc}.

If the physical layer employs fixed-length coding schemes without feedback, the \gls{urllc} design problem can be tackled using the nonasymptotic information-theoretic bounds developed in~\cite{polyanskiy10-05a} (see, e.g.,~\cite{yang14-07c, durisi16-02a, collins19-01, Ostman19-02}).
Such bounds allow one to assess for example how much frequency and spatial diversity should be exploited to achieve a target reliability for a given latency requirement.

Less is known in the \gls{vlsf} case.
The nonasymptotic achievability bound provided in~\cite[Thm.~3]{polyanskiy11-08a}, shows that, for a fixed reliability target, the  use of variable-length codes combined with stop feedback, allows one to approach capacity much faster in the (average) blocklength, compared to the scenario in which fixed-blocklength codes with no feedback are used.
However, the achievability bound given in~\cite[Thm.~3]{polyanskiy11-08a} pertains to the setup in which the acknowledgment sent on the feedback channel is assumed instantaneous and error-free.
As argued in, e.g.,~\cite{bennis18-10,Shariatmadari18-06}, these two assumptions are not suitable for \gls{urllc}.
The purpose of this paper is to generalize the analysis in~\cite[Thm.~3]{polyanskiy11-08a} and determine if \gls{vlsf} codes remain superior to fixed-blocklength no-feedback codes once the feedback delay and the presence of noise in the feedback link, which causes unreliable acknowledgments, are accounted for.

\paragraph*{Contributions}
Assuming arbitrary noisy forward and feedback channels, we obtain an upper bound on the error probability achievable using \gls{vlsf} coding schemes, for a fixed size of the information payload, and a given constraint on the maximum latency and on the average service time.
This last quantity is defined as the average time it takes the transmitter to process an information packet.
Our bound pertains to the setup in which there exists a constraint on the maximum number of transmission rounds (which is imposed by the latency requirement), after which an error is declared at the receiver.
Also, our analysis accounts for the presence of unreliable acknowledgments and of undetected error events, which occur whenever the decoder terminates transmission with an ACK, but its decision is erroneous.

The impact of unreliable acknowledgments can be mitigated through coding on the feedback channel.
However, this comes at a cost in terms of a feedback delay that is captured by our analysis.
Undetected errors are typically neglected in the analysis of  \gls{harq} protocols.
We argue that this simplifying assumption is unsuitable for the analysis of \gls{urllc} systems.
In practical systems, a \gls{crc} is typically used to detect errors at the receiver~\cite[Ch. 6.4]{dahlman11-a}.
Obviously, the longer the \gls{crc}, the lower the undetected error probability.
\markblue{However, for a given latency requirement, increasing the length of the \gls{crc} results in a reduction of the rate of the inner channel code.}
Hence, there is a fundamental trade-off that needs to be characterized for the optimal design of \gls{urllc} systems.
Our analysis, which relies on a threshold-based decoding rule that allows one to trade between reduction of service time and reduction of undetected error probability, sheds lights on this trade-off.

Focusing on the the \gls{urllc} regime where both reliability and latency requirements are stringent and the information payload is typically small, we use our bounds to \markblue{analyze} the performance of \gls{vlsf} coding schemes operating over practically relevant wireless channels.
Specifically, we consider transmissions over
\begin{inparaenum}[i)]
 \item a \gls{biawgn} channel and
 \item a block-memoryless Rayleigh fading channel, for the \markblue{practically relevant} setup in which  pilot symbols are used to estimate the channel coefficients, and the receiver is equipped with a mismatched decoder that treats the channel estimate as perfect.
\end{inparaenum}
In both cases, \markblue{our bound suggests} that the presence of noise on the feedback link causes a fundamental degradation in the performance of \gls{vlsf} codes.
\ju{For example, for the case of the \gls{biawgn} channel, when the size of the information payload is $30$ bits, the maximum latency constraint is $400$ channel uses, the packet error probability target is $10^{-5}$, and both the forward and the feedback channel operate at an SNR of $0\dB$, \markblue{our bounds results in an average-service-time estimate} of $106.6$ channel uses if the feedback link is assumed noiseless.
This value increases to $141$ channel uses if noise on the feedback link is accounted for.
For such a scenario, our bound suggests that the performance of  \gls{vlsf} coding schemes is inferior to that of a fixed-blocklength coding scheme without feedback, which has a (deterministic) service time of $130$ channel uses.}

\paragraph*{Prior Art}
To put our contribution into perspective, we survey next prior art on the analysis of the performance of point-to-point communication schemes with feedback.
Our review will focus on nonasymptotic results; hence, the vast literature that uses classical asymptotic information-theoretic metrics such as mutual information to characterize the performance of such systems will not be covered, since the results obtained following this approach are often not relevant for the design of \gls{urllc} systems.

One way to provide nonasymptotic performance analyses of both fixed-length and variable-length coding schemes is through the characterization of the reliability function, which determines the speed at which the error probability vanishes as a function of the blocklength for a fixed communication rate.
When no feedback is available, the reliability function in the fixed-blocklength case is known to be no larger than the so-called sphere-packing bound~\cite{shannon67-a}.
Furthermore, the sphere-packing bound is achievable for all rates between the critical rate and capacity~\cite{gallager68a}.
For the case of symmetric \glspl{dmc}, it is known that the reliability function does not exceed the sphere-packing bound even when noiseless full feedback\footnote{By full feedback we mean that the transmitter has perfect causal knowledge of the channel outputs at the receiver.} is available~\cite{dobrushin62-a}.
However, as recently shown in~\cite{burnashev10-01} for the case of \glspl{bsc}, when the rate is below the critical rate, the availability of full feedback, even when noisy, allows one to operate above the best known lower bound on the reliability function for the no-feedback case.

The use of variable-length codes together with the availability of noiseless full feedback results in a much improved reliability function compared to the fixed-length full-feedback case~\cite{burnashev76-12a}.
Such a significant improvement can be also observed in the moderate-deviation regime, in which the rate tends to capacity and the error probability tends to zero as the blocklength tends to infinity~\cite{Truong19}.

The reliability function for the variable-length full-feedback case remains above the sphere-packing bound even when the full feedback is noisy~\cite{Draper08-04}.
As noted in~\cite{Draper08-04}, noise in the feedback link may cause synchronization errors that need to be accounted for in the analysis.
This can be done, for example, by using the framework for tracking stopping times through noisy observations put forward in~\cite{Niesen09}.

Variable-length codes combined with stop feedback rather than full feedback were analyzed in~\cite{forney-jr68-03a,telatar92-05a} where it is shown that in the noiseless case, the reliability function exceeds the sphere-packing bound.

For the fixed-blocklength no-feedback case, the recent work by Polyanskiy \emph{et al.}~\cite{polyanskiy10-05a} has renewed interest in determining nonasymptotic bounds on the minimum error probability that are tighter than the ones obtainable through a reliability-function analysis.
The bounds provided in~\cite{polyanskiy10-05a} allow one to obtain tight performance characterizations also in the so-called normal regime, where the transmission rate is close to capacity.
Furthermore, in this regime the bounds, \markblue{which typically do not admit a closed-form expression}, can be efficiently approximated using a compact expression commonly referred to as normal approximation.

As shown  in, e.g.,~\cite{polyanskiy10-05a,wu11-12}, the normal approximation can be used to analyze the non-asymptotic performance of simple \gls{arq} schemes.
In particular, the authors of~\cite{wu11-12} studied a good-put maximization problem for the case in which simple \gls{arq} is used over a Rayleigh block-fading channel and the stop feedback is noisy.
By leveraging the normal approximation, and by employing a simple model for imperfect error detection, they determine the blocklength required on both the forward and the feedback channel in order to maximize the good-put.

The use  of the normal approximation for the analysis of general \gls{harq} schemes (see, e.g., \cite{Avranas18,Makki18-11}) is not entirely satisfactory from a theoretical viewpoint.
Indeed, such an approach does not guarantee the existence of a single \markblue{mother code} that, when shortened to an arbitrary blocklength, achieves the error probability predicted by the normal approximation.

A rigorous analysis of the error probability achievable using general VLSF coding schemes was undertaken by Polyanskiy \emph{et al.} in~\cite{polyanskiy11-08a}.
Specifically, they provided in~\cite[Thm.~3]{polyanskiy11-08a} a nonasymptotic upper bound on the minimum error probability achievable with \gls{vlsf} coding schemes, which reveals that the maximum coding rate achievable with \gls{vlsf} coding schemes converges faster to capacity as the blocklength increases, compared to the fixed-blocklength, no-feedback case.
The nonasymptotic upper bound in~\cite[Thm.~3]{polyanskiy11-08a} relies on a \gls{vlsf} coding scheme in which the decoder computes the accumulated information density corresponding to each possible codeword and sends a stop signal whenever one of the accumulated information densities exceeds a threshold.
In this scheme, the number of transmission rounds is unlimited, and an ACK/NACK bit is fed back over a noiseless channel after the reception of each symbol.

An extension of this upper bound to the case in which the number of transmission rounds is finite and the feedback bit is transmitted only after a block of symbols can be found in~\cite{kim15-04, williamson15-07a}.
Furthermore, feedback delay is accounted for in~\cite{ostman18-12}.
Finally, adaptations of~\cite[Thm.~3]{polyanskiy11-08a} to the case of random packet arrivals and to the case of common-message transmission over a broadcast channel are provided in~\cite{rahul19-04} and~\cite{Trillinsgaard18-12}, respectively.

A different approach to bounding the error probability of \gls{vlsf} coding schemes is presented in~\cite{Malkamaki00-09}.
There, a random coding bound is obtained for the setup in which a low-rate inner code, used to provide incremental redundancy, is combined with a high-rate \gls{crc}.
The analysis provided in~\cite{Malkamaki00-09}, which relies on an error-exponent bound, pertains to the transmission of binary antipodal coded symbols over a block-fading channel and accounts for the presence of noise in the feedback link.
The authors, however, assume for simplicity that a NACK cannot be interpreted as ACK by the transmitter---a simplification we dispose with in our analysis.
Finally, they present a comparison between the bounds and the performance of actual \gls{vlsf} coding schemes relying on convolutional codes.

 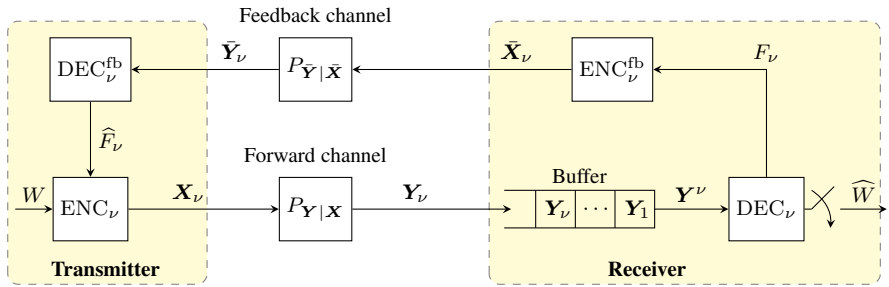
\begin{figure*}[t]
     \centering
         \begin{tikzpicture}[scale=1]
\tikzstyle{every node}=[font=\scriptsize]

\tikzset{
block/.style = {draw, fill=white, rectangle, minimum height=3em, minimum width=3em},
bufferblock/.style = {draw, fill=white, rectangle, minimum height=2cm, minimum width=1cm},
tmp/.style  = {coordinate},
sum/.style= {draw, fill=white, circle, node distance=1cm},
input/.style = {coordinate},
output/.style= {coordinate},
pinstyle/.style = {pin edge={to-,thin,black}},
queuei/.pic={
  \draw (0,0) -- ++(2cm,0) -- ++(0,-0.5cm) -- ++(-2cm,0);
   \foreach \c [count =\x from 1] in {$\randvecy_1$,$\cdots$,$\randvecy_{\nu}$}
   {
    \draw ([xshift=-\x*15pt]2cm,0) -- ++(0,-0.5cm) {};
    \node[text centered] at ([xshift=-\x*15pt]2.3cm,-0.25cm) {\c};
   }
   \node[above] at (1cm,0) {Buffer};
  },
  bsc/.pic={
     \node[circle,fill=blue,inner sep=1pt,minimum size=1pt] (A1) at (0,0.5) {} ;
      \node[circle,fill=blue,inner sep=1pt,minimum size=1pt] (A2)  at (0,0) {} ;
     \node[circle,fill=blue,inner sep=1pt,minimum size=1pt] (B1) at (0.5,0.5) {} ;
      \node[circle,fill=blue,inner sep=1pt,minimum size=1pt] (B2)  at (0.5,0) {};
      \draw[-]  (A1)--(B1) node [midway, above, sloped]  {};
      \draw[-]  (A1)--(B2) node at (1,2.4) {};
      \draw[-]  (A2)--(B1) node  at (1,0.6) {};
      \draw[-]  (A2)--(B2) node [midway, below, sloped]  {};
  }
}

	 \path[fill=yellow!20,rounded corners, draw=black!50, dashed] (-0.1,-1) rectangle (2.55,2.5) ;
	 \node at (1.2,-0.8) {\textbf{Transmitter}} ;
	 \path[fill=yellow!20,rounded corners, draw=black!50, dashed] (6.3,-1) rectangle (11.5,2.5) ;
	\node   at (8.4,-0.8) {\textbf{Receiver}};

    \node [input, name=rinput] (rinput) {};
	\node [block, right of=rinput] (enc) {$\mathrm{ENC}_\nu$};
    \node [block, right = 2cm of enc] (channel) {$P_{\randvecy \given \randvecx }$};
    \node[above =0.1 cm of channel] {Forward channel};

    \path  (6.5cm,0.25cm) pic {queuei=1};
	\node [block, right = 5cm of channel] (dec) {$\mathrm{DEC}_\nu$};

    \node [block, above left  = 1cm and 1 cm  of dec] (enc2) {$\mathrm{ENC}^{\mathrm{fb}}_\nu$};
    \node [block, left = 2.375cm of enc2, above = 1cm of channel] (feedchannel) {$P_{\bar{\randvecy} \given \bar{\randvecx}}$};
    \node[above =0.1 cm of feedchannel] {Feedback channel};
	\node [above right = 0.5cm  and 1cm of dec] (feedback) {};
	\node [block, above = 1 cm of enc] (dec2) {$\mathrm{DEC}^{\mathrm{fb}}_\nu$};

	\draw[->] (rinput) -- node[above] {$W$}  (enc);
    \draw[->] (enc) -- node[above, pos=0.4] {$\randvecx_\nu$} (channel);
 	\draw[->] (channel) -- node[above, pos=0.4] {$\randvecy_\nu$} +(2.6cm,0);
 	\draw[->] (8.5cm, 0cm) --  node[above] {$\randvecy^\nu$} (dec);

 	 \draw[->](dec) |- node [above] {$F_\nu $} (enc2);
	\draw[->](enc2) -- node [above, pos=0.25] {$\bar{\randvecx}_\nu $ } (feedchannel);
	\draw[->](feedchannel) -- node [above, pos=0.3] {$\bar{\randvecy}_\nu $ }  (dec2);
	\draw[->](dec2) -- node[right]{$\widehat{F}_\nu$}  (enc);

 	 \draw[-] (dec) -- ($(dec)+(0.6cm,0)$);
 	 \draw[-] ($(dec)+(0.6cm,0)$)-- ($(dec)+(0.9cm, 0.3cm)$);
 	 \draw[->]  ($(dec)+(0.98cm, 0)$) -- node[above]{$\widehat{W}$} ($(dec)+(1.6cm, 0cm)$);

 	 \draw[->] ($(dec)+(0.6cm, 0.25cm)$) to [bend left ] ($(dec)+(0.85cm, -0.25cm)$);

\end{tikzpicture}
     \caption{Round $\nu$ of the \gls{vlsf} transmission scheme. Here, $W$ denotes the information message, $\widehat{W}$ is its estimate at the receiver, $F_\nu$ is the feedback bit computed at the receiver, and $\widehat{F}_\nu$ is its estimate at the transmitter.}
     \label{fig:sys_mod}
 \end{figure*}

  \paragraph*{Notation}
 Upper case letters are used to denote random vectors, e.g., $\randvecx$ and their realizations are written in lower case, e.g., $\vecx$.
 The probability distribution of $\randvecx$ is written as $P_{\randvecx}$.
 We use superscripts to denote the concatenation of vectors of equal size, e.g., $\randvecx^\nu=\lrho{\randvecx_1, \dots, \randvecx_\nu}$.
 The distribution of a real Gaussian random variable is denoted by $\mathcal{N}\lro{\mu,\sigma^2}$ and the distribution of a complex proper Gaussian random variable is denoted by $\mathcal{CN}\lro{\mu, \sigma^2}$.
 Here, $\mu$ and $\sigma^2$ are the mean and the variance of the random variable, respectively.
 The Radon-Nikodym derivative of a distribution $P_1$ with respect to a distribution $P_2$, where $P_1$ is absolutely continuous with respect to $P_2$, is denoted by $\frac{\mathrm{d}P_1}{\mathrm{d}P_2}$.
 Finally, \Ex{}{\cdot} is the expectation operator, $\prob{\cdot}$ is used for probabilities, $\mathbb{1}\{\cdot\}$ denotes the indicator function, $Q\lro{\cdot}$ stands for the Gaussian Q-function, \markblue{and $I(P_X,P_{Y\given X})$ denotes the mutual information between the random variables $X$ and $Y$, whose joint distribution is $P_XP_{Y\given X}$}.

  \section{System Model}\label{sec:sys_mod}

  We consider a point-to-point communication system in which information is transmitted using the~\gls{vlsf} coding scheme depicted in Fig.~\ref{fig:sys_mod}.
  Specifically, transmission occurs over a variable number of rounds.
  Each round is divided into a data phase and a feedback phase, not necessarily of equal duration.
  Throughout, we assume that the number of transmission rounds does not exceed the integer $\nmax<\infty$.

  In the data phase, a segment (spanning $n$ channel uses) of the codeword associated to the current information message is sent to the receiver over the forward channel.
  This channel is modeled as a sequence of conditional probability kernels $\{P_{\randvecy_\nu\given \randvecy^{\nu-1},\randvecx^{\nu}}\}_{\nu=1}^{\nmax}$, where the random vectors $\randvecy_\nu$ and $\randvecx_\nu$, $\nu=1,\dots,\nmax$, take values from the sets $\setY^n$ and $\setX^n$, respectively.

  \markblue{For analytical tractability, we assume that the channel is block-wise stationary and memoryless, i.e.,
  \begin{equation}\label{eq:forward_channel}
    P_{{\randvecy_\nu}\given {\randvecy}^{\nu-1},{\randvecx}^{\nu}}({\vecy}_\nu\given {\vecy}^{\nu-1},{\vecx}^{\nu})=P_{{\randvecy}\given{\randvecx}}({\vecy}_\nu\given{\vecx}_\nu).
  \end{equation}
  }

  At the end of each data phase, the receiver decides whether to perform decoding based on the channel outputs received that far, or to request an additional transmission.
  The outcome of this decision---a single bit of information conveying the message ``stop'', which we denote by $\ack$ or ``continue'', which we denote by $\nack$, is transmitted in the feedback phase over the feedback channel using $\nf$ channel uses.\footnote{The symbol $\ack$ corresponds to an ACK, whereas the symbol $\nack$ corresponds to a NACK.}
  We model the feedback channel as a sequence of conditional probability kernels $\{P_{\bar{\randvecy}_\nu\given \bar{\randvecy}^{\nu-1},\bar{\randvecx}^{\nu}}\}_{\nu=1}^{\nmax}$, where the random vectors $\bar{\randvecy}_\nu$ and $
  \bar{\randvecx}_\nu$, $\nu=1,\dots,\nmax$, take values from the sets $\bar{\setY}^{\nf}$ and $\bar{\setX}^{\nf}$, respectively.
  \markblue{As for the forward channel,} we assume for simplicity that the feedback channel is block-wise stationary and memoryless, i.e.,
  \begin{equation}\label{eq:feedback_channel}
    P_{\bar{\randvecy_\nu}\given \bar{\randvecy}^{\nu-1},\bar{\randvecx}^{\nu}}(\bar{\vecy}_\nu\given \bar{\vecy}^{\nu-1},\bar{\vecx}^{\nu})=P_{\bar{\randvecy}\given\bar{\randvecx}}(\bar{\vecy}_\nu\given\bar{\vecx}_\nu).
  \end{equation}
  This enables a single encoder/decoder pair to be used on the feedback channel over consecutive transmission rounds.

  Upon observing the output of the feedback channel, the transmitter decides whether $\ack$ or $\nack$ was sent.
  This implies that the feedback channel, \markblue{together with the signaling scheme just described}, can be viewed as a binary asymmetric channel, with crossover probabilities $\efa=\prob{\ack\rightarrow \nack}$ and $\efb=\prob{\nack\rightarrow \ack}$, which depend both on $\nf$ and on the encoder-decoder pair used to transmit the binary message over the feedback channel.


  Some remarks on our setup are in order.
  We allow for $\efa\neq \efb$ since the $\ack \rightarrow \nack$ and the $\nack \rightarrow \ack$ events have a different impact on performance.
  Indeed, the $\nack \rightarrow \ack$ event causes the premature interruption of the transmission of the current message.
  We assume, somewhat pessimistically, that this always results in an error at the decoder.
  This error needs to be handled by higher layers, often causing a violation of the latency requirement.
  On the contrary, the $\ack \rightarrow \nack$ event triggers an unnecessary additional transmission round, which causes only a moderate increase in service time.
  Assuming that these two error events on the feedback channel have different probability is in agreement with current wireless standards, where one typically imposes that $\efb \ll \efa$.
  For example, in \gls{lte}, we typically have $\efa=10^{-2}$ and $\efb \in [10^{-4},10^{-3}]$~\cite[Ch. 10.4.2]{dahlman11-a}.

  Note that an error on the feedback channel may result in the transmitter and the receiver falling out of synchronization, i.e., operating on different messages.
  \markblue{To prevent this, we assume that each codeword segment contains a binary flag specifying whether the segment is the first one of a new information-message transmission or not.
  Through coding, one can ensure that this flag is transmitted with a sufficiently high reliability, to avoid synchronization issues.
  Throughout the paper, we assume for simplicity that this flag is always received correctly at the decoder.
  From a modeling perspective, this is equivalent to assuming that the noisy estimate of the feedback bit produced at the transmitter is known to the receiver.
  In Section~\ref{sec:conclusion}, we discuss how to generalize our analysis to account for errors in the transmission of this flag.}

\begin{figure*}[t]
    \centering
    \begin{subfigure}[t]{0.3\textwidth}
      \centering
%
%
%
%
%
%

\def \bufferlength{0.3}
\def \bufferwidth{0.3}
\def \buffery{-0.3}
\def \pktwidth{1}
\def \msgwidth{0.2}
\def \FBpktwidth{0.2}
\def \roundwidth{1.2}
\def \datapktHeight{0.2}
\def \fbpktHeight{1}
\def \ctrlpktHeight{0.6}

\begin{tikzpicture}[thick,scale=1.4, ]
\tikzstyle{every node}=[font=\scriptsize]


\tikzset{
  cross/.style={thick,cross out, draw=red, minimum size=2*(#1-\pgflinewidth), inner sep=0pt, outer sep=0pt},
  cross/.default={3pt},
  cross/.style={very thick,cross out, draw=red, minimum size=2*(#1-\pgflinewidth), inner sep=1pt, outer sep=1pt},
  cross/.default={3pt}
}

\draw[->] (0,\datapktHeight-1) -- +(2.8,0) node[anchor=north, yshift=-0.1cm ] {time};


\foreach \x in {0,...,2} {
  \draw[-,>=stealth, thick] (\x*\pktwidth+\x*\FBpktwidth,\datapktHeight-1-0.1) -- +(0,0.2);
}

\draw[draw=black] (0,\datapktHeight) rectangle ++(\pktwidth,0.3) node[pos=.5]{$\mathrm{ENC}_1(5)$};
\draw[draw=black] (\pktwidth+\FBpktwidth,\datapktHeight) rectangle ++(\pktwidth,0.3) node[pos=.5]{$\mathrm{ENC}_2(5)$};


\draw[draw=black, fill=green!40!white] (\pktwidth,\fbpktHeight) rectangle ++(\FBpktwidth,0.3)node[pos=.5]{$\mathrm{s}$};
\draw[draw=black, fill=green!40!white] (2*\pktwidth+\FBpktwidth,\fbpktHeight) rectangle ++(\FBpktwidth,0.3)node[pos=.5]{$\mathrm{s}$};

\draw[draw=black, fill=red!40!white] (\pktwidth,\ctrlpktHeight) rectangle ++(\FBpktwidth,0.3)node[pos=.5]{$\mathrm{c}$};
\draw[draw=black, fill=green!40!white] (2*\pktwidth+\FBpktwidth,\ctrlpktHeight) rectangle ++(\FBpktwidth,0.3)node[pos=.5]{$\mathrm{s}$};

\node[] at (0,\datapktHeight+0.15-0.4) { $5$};
\node[] at (\roundwidth,\datapktHeight+0.15-0.4) { $5$};

\node[] at (\pktwidth,\datapktHeight+0.15-0.8) { $9$};

\node[anchor = west] at (-0.5,\datapktHeight+0.15-0.4)  { $W$};
\node[anchor = west] at (-0.5,\datapktHeight+0.15-0.8)  { $\widehat{W}$};
\node[anchor=west] at (-0.5,\datapktHeight+0.15) (a) { $\randvecx_\nu$};
\node[anchor=west] at (-0.5,\fbpktHeight+0.15) (a) { $F_\nu$};
\node[anchor=west] at (-0.5, \ctrlpktHeight+0.15) { $\widehat{F}_\nu$};


%
%
\end{tikzpicture}

      \caption{\ju{The message $W=5$ is incorrectly decoded as $\widehat{W}=9$ after the first round; furthermore, an $\ack\rightarrow \nack$ event causes the retransmission of message $W=5$.
      Note: an $\ack \rightarrow \nack$ event causes the retransmission of $W$, even when there is no decoding error.}}
      \label{fig:errors_a}
    \end{subfigure}
    \hspace{0.1cm}
    \begin{subfigure}[t]{0.3\textwidth}
      \centering
%
%
%
%
%
%

\def \bufferlength{0.3}
\def \bufferwidth{0.3}
\def \buffery{-0.3}
\def \pktwidth{1}
\def \msgwidth{0.2}
\def \FBpktwidth{0.2}
\def \roundwidth{1.2}
\def \datapktHeight{0.2}
\def \fbpktHeight{1}
\def \ctrlpktHeight{0.6}

\begin{tikzpicture}[thick,scale=1.4, ]
\tikzstyle{every node}=[font=\scriptsize]


\tikzset{
  cross/.style={thick,cross out, draw=red, minimum size=2*(#1-\pgflinewidth), inner sep=0pt, outer sep=0pt},
  cross/.default={3pt},
  cross/.style={very thick,cross out, draw=red, minimum size=2*(#1-\pgflinewidth), inner sep=1pt, outer sep=1pt},
  cross/.default={3pt}
}

\draw[->] (0,\datapktHeight-1) -- +(2.8,0) node[anchor=north, yshift=-0.1cm ] {time};


\foreach \x in {0,...,2} {
  \draw[-,>=stealth, thick] (\x*\pktwidth+\x*\FBpktwidth,\datapktHeight-1-0.1) -- +(0,0.2);
}

\draw[draw=black] (0,\datapktHeight) rectangle ++(\pktwidth,0.3) node[pos=.5]{$\mathrm{ENC}_1(5)$};
\draw[draw=black] (\pktwidth+\FBpktwidth,\datapktHeight) rectangle ++(\pktwidth,0.3) node[pos=.5]{$\mathrm{ENC}_1(7)$};


\draw[draw=black, fill=red!40!white] (\pktwidth,\fbpktHeight) rectangle ++(\FBpktwidth,0.3)node[pos=.5]{$\mathrm{c}$};

\draw[draw=black, fill=green!40!white] (\pktwidth,\ctrlpktHeight) rectangle ++(\FBpktwidth,0.3)node[pos=.5]{$\mathrm{s}$};

\node[] at (0,\datapktHeight+0.15-0.4) { $5$};
\node[] at (\roundwidth,\datapktHeight+0.15-0.4) { $7$};

\node[] at (\roundwidth+\pktwidth,\datapktHeight+0.15-0.8) { $\mathrm{e}$};

\node[anchor = west] at (-0.5,\datapktHeight+0.15-0.4)  { $W$};
\node[anchor = west] at (-0.5,\datapktHeight+0.15-0.8)  { $\widehat{W}$};
\node[anchor=west] at (-0.5,\datapktHeight+0.15) (a) { $\randvecx_\nu$};
\node[anchor=west] at (-0.5,\fbpktHeight+0.15) (a) { $F_\nu$};
\node[anchor=west] at (-0.5, \ctrlpktHeight+0.15) { $\widehat{F}_\nu$};



%
%
\end{tikzpicture}

      \caption{A $\nack\rightarrow \ack$ event causes the transmitter \markblue{to remove message $W=5$ from its buffer, and move to the new message} $W=7$ after the first round. The receiver observes a packet out of sequence and declares an error.}
      \label{fig:errors_b}
    \end{subfigure}
    \hspace{0.1cm}
    \begin{subfigure}[t]{0.3\textwidth}
      \centering
%
%
%
%
%
%

\def \bufferlength{0.3}
\def \bufferwidth{0.3}
\def \buffery{-0.3}
\def \pktwidth{1}
\def \msgwidth{0.2}
\def \FBpktwidth{0.2}
\def \roundwidth{1.2}
\def \datapktHeight{0.2}
\def \fbpktHeight{1}
\def \ctrlpktHeight{0.6}

\begin{tikzpicture}[thick,scale=1.4, ]
\tikzstyle{every node}=[font=\scriptsize]


\tikzset{
  cross/.style={thick,cross out, draw=red, minimum size=2*(#1-\pgflinewidth), inner sep=0pt, outer sep=0pt},
  cross/.default={3pt},
  cross/.style={very thick,cross out, draw=red, minimum size=2*(#1-\pgflinewidth), inner sep=1pt, outer sep=1pt},
  cross/.default={3pt}
}

\draw[->] (0,\datapktHeight-1) -- +(3.8,0) node[anchor=north, yshift=-0.1cm ] {time};


\foreach \x in {0,...,3} {
  \draw[-,>=stealth, thick] (\x*\pktwidth+\x*\FBpktwidth,\datapktHeight-1-0.1) -- +(0,0.2);
}

\draw[draw=black] (0,\datapktHeight) rectangle ++(\pktwidth,0.3) node[pos=.5]{$\mathrm{ENC}_1(5)$};
\draw[draw=black] (\pktwidth+\FBpktwidth,\datapktHeight) rectangle ++(\pktwidth,0.3) node[pos=.5]{$\mathrm{ENC}_2(5)$};
\draw[draw=black] (2*\pktwidth+2*\FBpktwidth,\datapktHeight) rectangle ++(\pktwidth,0.3) node[pos=.5]{$\mathrm{ENC}_3(5)$};


\draw[draw=black, fill=red!40!white] (\pktwidth,\fbpktHeight) rectangle ++(\FBpktwidth,0.3)node[pos=.5]{$\mathrm{c}$};
\draw[draw=black, fill=red!40!white] (2*\pktwidth+\FBpktwidth,\fbpktHeight) rectangle ++(\FBpktwidth,0.3)node[pos=.5]{$\mathrm{c}$};

\draw[draw=black, fill=red!40!white] (\pktwidth,\ctrlpktHeight) rectangle ++(\FBpktwidth,0.3)node[pos=.5]{$\mathrm{c}$};
\draw[draw=black, fill=red!40!white] (2*\pktwidth+\FBpktwidth,\ctrlpktHeight) rectangle ++(\FBpktwidth,0.3)node[pos=.5]{$\mathrm{c}$};

\node[anchor = west] at (-0.5,\datapktHeight+0.15-0.4)  { $W$};
\node[anchor = west] at (-0.5,\datapktHeight+0.15-0.8)  { $\widehat{W}$};
\node[anchor=west] at (-0.5,\datapktHeight+0.15) (a) { $\randvecx_\nu$};
\node[anchor=west] at (-0.5,\fbpktHeight+0.15) (a) { $F_\nu$};
\node[anchor=west] at (-0.5, \ctrlpktHeight+0.15) { $\widehat{F}_\nu$};

\node[] at (0,\datapktHeight+0.15-0.4) { $5$};
\node[] at (\roundwidth,\datapktHeight+0.15-0.4) { $5$};
\node[] at (2*\roundwidth,\datapktHeight+0.15-0.4) { $5$};

\node[] at (2*\roundwidth+\pktwidth,\datapktHeight+0.15-0.8) { $\mathrm{e}$};

\end{tikzpicture}

      \caption{The receiver declares an error after ${\nmax=3}$ unsuccessful transmission rounds.}
      \label{fig:errors_c}
    \end{subfigure}
  \caption{Example of the three types of errors for $\nmax=3$.
  Here, $W$ is the transmitted message and $\widehat{W}$ is the estimate of $W$ at the receiver.
Furthermore, $F_\nu$ is the feedback bit generated by the receiver in round $\nu$, and $\widehat{F}_\nu$ is its estimate at the transmitter.}
  \label{fig:errors}
  \end{figure*}
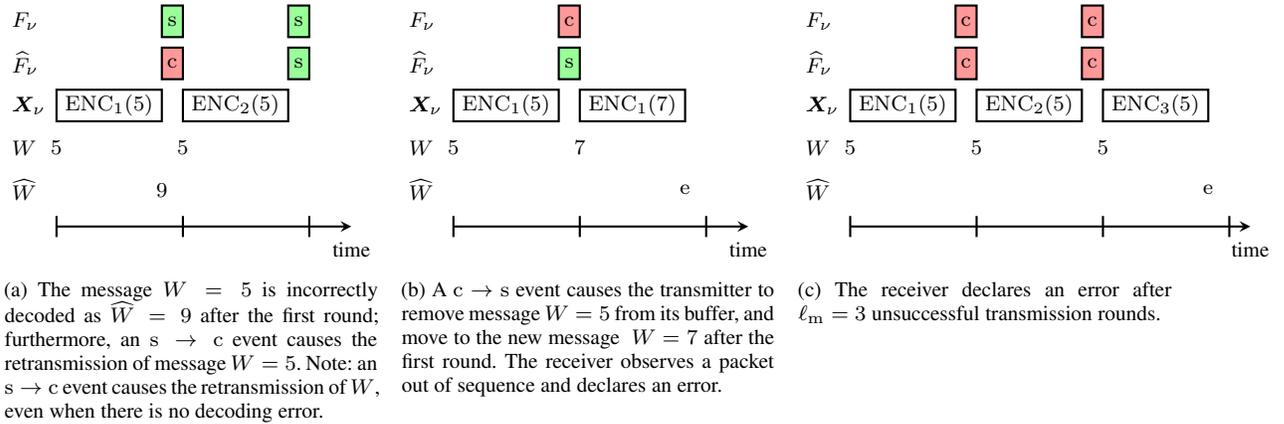

  To summarize, in our setup, a transmission error occurs if
  \begin{itemize}
    \item The receiver decides to perform decoding but produces the wrong codeword estimate---an event typically referred to as \emph{undetected error}.
    This event is shown in Fig. \ref{fig:errors_a} along with an $\ack\rightarrow \nack$ event, which does not cause an error, but increases the service time.

    \item A $ \nack \rightarrow \ack $ event occurs on the feedback channel, see Fig. \ref{fig:errors_b}.

    \item The receiver is not able to perform decoding within the available $\nmax$ rounds, see Fig. \ref{fig:errors_c}.
  \end{itemize}
  In the last two cases, the decoder declares an erasure, which we denote by the symbol $\mathrm{e}$.

  \subsection{Definition of a \gls{vlsf} Code}\label{sec:vlsf_code}
  Before providing a formal definition of a \gls{vlsf} coding scheme for the noisy feedback case, we introduce some additional notation.
  We let  $F_\nu \in \{\ack,\nack\}$ be the feedback bit generated by the receiver in round $\nu=1,\dots,\nmax$ and  $\widehat{F}_\nu \in \{\ack,\nack\}$ its estimate at the transmitter.

  Note that in the presence of errors on the feedback link, the number of rounds after which the receiver produces an estimate of the transmitted message (or declares an erasure) does not necessarily coincide with the number of transmission rounds (see Fig.~\ref{fig:errors_a}).

  As a consequence, the average service time at the transmitter, which is the average number of transmission rounds after which the current message is removed from the buffer at the transmitter, does not generally coincide with the average latency at the receiver, which is the average number of transmission rounds needed by the receiver to produce a message estimate or to declare an erasure.

  From a system-level perspective, the average service time at the transmitter is relevant in full-buffer scenarios, where one is interested in maximizing the long-term throughput.
  Indeed, according to the renewal-reward theorem, this quantity is given by the ratio between the number of information bits per message and the average service time at the transmitter.
  Minimizing the average service time at the transmitter is also of interest whenever an objective is to minimize the average energy consumption.
  \ju{Hence, achieving a small service time is of interest also in sporadic transmissions.}

  Throughout the paper, we shall focus mainly on the case in which the average service time at the transmitter is the metric of interest.
  However, we will also discuss how to adapt our analysis to the case in which the metric of interest is the average latency at the receiver.

  The definition of a \gls{vlsf} coding scheme provided below is an adaptation to the noisy feedback case of the definition of a  \gls{vlsf} coding scheme given in~\cite{polyanskiy11-08a}.
  \begin{dfn}\label{def:code}
    \ju{An $\lro{\ell\sub{a}, M, \epsilon, \nmax,n,\nf}$-\gls{vlsf} coding scheme where $M$, $\nmax$, $n$, and $\nf$ are positive integers, $\ell\sub{a}$ is a nonnegative real number, and $\epsilon \in (0,1)$, consists of:}
    \begin{itemize}
      \item A random variable $U$ defined on a set $\setU$ of cardinality $\abs{\setU}\leq 2$  that is revealed to both the transmitter and the receiver before the start of the transmission.
      This random variable acts as common randomness and allows for the use of \markblue{randomized coding strategies}.
      \item A sequence of $\nmax$ encoders for the forward channel $f_\nu: \setU \times \lrbo{1,\dotsc,M} \rightarrow \setX^n$, $\nu=1,\dots,\nmax$, defining the forward-channel input
      \begin{equation}\label{eq:channel_input}
        \randvecx_\nu=f_\nu(U,W)
      \end{equation}
      for a given message $W$, which we assume to be uniformly distributed over $\lrbo{1,\dotsc,M}$.

      \item A sequence of $\nmax$ decoders for the forward channel $g_\nu: \setU \times \mathcal{Y}^{n\nu}  \rightarrow \lrbo{1,\dotsc,M}$, $\nu=1,\dots,\nmax$, providing an estimate $g_\nu(U,\randvecy^{\nu})$ of the message $W$.

      \item A sequence of binary random variables $F_\nu \in \{\ack,\nack\}$, $\nu=1,\dots,\nmax$, each being the outcome of the evaluation of a stopping rule defined on the filtration $\sigma(U,\randvecy_1,\dots \randvecy_\nu)$.
      These random variables are the binary messages transmitted by the receiver  on the feedback channel.

      \item An encoder for the feedback channel $\bar{f}: \{\ack,\nack\} \rightarrow \bar{\setX}^{\nf}$ defining the feedback-channel input $\bar{\randvecx}_\nu=\bar{f}(F_\nu)$ at round $\nu=1,\dots,\nmax$.

      \item A decoder for the feedback channel $\bar{g}: \bar{\setY}^{\nf}\rightarrow \{\ack,\nack\}$ that produces the estimate $\widehat{F}_\nu=\bar{g}(\bar{\randvecy}_\nu)$ at round $\nu$.

      \item Two stopping times, one at the transmitter $\ttx$ and one at the receiver $\trx$, and a message estimate $\widehat{W} \in \{1,\dots, M\} \cup \mathrm{e}$, all defined through the procedure detailed in Algorithm~\ref{alg:algorithm}.
      The stopping time $\ttx$ satisfies the average service-time constraint
      \begin{equation}\label{eq:avg_latency}
       \Ex{}{\ttx}  \leq \ell\sub{a}
      \end{equation}
      and the message estimate $\widehat{W}$ satisfies the error probability constraint
      \begin{equation}\label{eq:error}
        \prob{\widehat{W}\neq W }\leq \epsilon.
      \end{equation}
    \end{itemize}
  \end{dfn}

  \begin{algorithm}
  \caption{Procedure at the transmitter and the receiver to compute the message estimate~$\widehat{W}$, the transmitter stopping time~$\ttx$,  and the receiver stopping time~$\trx$.}\label{alg:algorithm}
  \begin{algorithmic}
  \Initialize{ $\ttx=\trx=\infty$;\, $F_{0} = \widehat{F}_{0} =\nack ;$}
  \For {$\nu = 1 \rightarrow \nmax$}
    \State Transmitter:
        \If{$\nu>1$}
          \State compute $\widehat{F}_{\nu-1}=\bar{g}(\bar{\randvecy}_{\nu-1})$
        \EndIf
        \If{$\widehat{F}_{\nu-1}=\nack$}
          \State transmit $f_\nu(U,W)$ over the forward channel
          \If{$\nu=\nmax$}
            \State set $\ttx=\nmax$
          \EndIf
        \Else
            \State set $\ttx=\nu-1$
        \EndIf
  %
    \State Receiver:

    \Switch{$(F_{\nu-1}, \widehat{F}_{\nu-1})$}
          \Case{$(\ack,\ack)$}
            \State STOP
          \EndCase
          \Case{$(\ack,\nack)$}
            \State  set $F_\nu=\ack$
          \EndCase
          \Case{$(\nack,\ack)$}
            \State set $\trx=\nu$, $\widehat{W}=\mathrm{e}$
            \State STOP
          \EndCase
          \Case{$(\nack,\nack)$}
          \State use stopping rule to compute $F_\nu$
          \If   {$F_\nu=\ack$}
            \State set $\widehat{W}=g_{\nu}(\randvecy^\nu,U)$ and $\trx=\nu$
          \EndIf
          \EndCase
      \EndSwitch
      \State \textbf{end switch}

         \If    {$\nu<\nmax$}
            \State send $\bar{f}(F_\nu)$ on the feedback channel
         \Else
            \If {$\trx=\infty$}
              \State set $\widehat{W}=\mathrm{e}$ and $\trx=\nmax$
            \EndIf
         \EndIf
  \EndFor
  \end{algorithmic}
  \end{algorithm}

  Some remarks are in order.
  Compared to the definition of \gls{vlsf} codes provided in~\cite{polyanskiy11-08a}, which involves a single stopping time at the receiver, our definition involves two stopping times, one at the transmitter and one at the receiver.
  This is needed to account for errors on the feedback link.
  Also, the decoder employs an erasure option, which is used if a $\nack\to\ack$ event occurs, or if the stopping rule is not triggered after~$\nmax$ rounds.
  Note that we measure the service time in transmission rounds.
  Each transmission round involves~$n$ channel uses on the forward channel and~$\nf$ channel uses on the feedback channel.

  Our definition can be readily adapted to the case in which the average latency at the receiver is the metric of interest.
  Indeed, it is sufficient to replace $\Ex{}{\ttx}$ in~\eqref{eq:avg_latency} with $\Ex{}{\trx}$.

  \markblue{The random variable $U$, which also appears in the definition of \gls{vlsf} codes provided in~\cite{polyanskiy11-08a}, enables the use of randomized coding strategies, which, as we shall see in the proof of our main result, are needed to obtain bounds on~\eqref{eq:avg_latency} and~\eqref{eq:error} through a random coding argument.
  }
  %

  \section{Main Result}\label{sec:main_result}
  We provide an achievability bound, i.e., an upper bound on the error probability achievable using \gls{vlsf} coding schemes defined according to Definition~\ref{def:code}, for a fixed number of messages $M$, a fixed average service time $\ell\sub{a}$, and a fixed latency requirement $\nmax$.

  Before presenting our bound, we characterize  the pairs $(\efa=\prob{\ack\rightarrow \nack}, \efb=\prob{\nack\rightarrow \ack})$ that are achievable for a given choice of the encoder for the feedback channel.

  \begin{lem}\label{lem:hypothesis-testing}
  For a given  $\nf$ and for a given encoder $\bar{f}: \{\ack,\nack\} \rightarrow \bar{\setX}^{\nf}$ for the feedback channel, all pairs $(\efa,\efb)$ in the convex hull of the union on the following two sets are achievable
  \markblue{
    \begin{IEEEeqnarray}{rCl}
      \bigcup_{\gamma\sub{f}\in \reals \cup \{\pm \infty\}}
      \left(
       \prob{
       \frac{\mathrm{d}P^{(\nack)}}{
       \mathrm{d}P^{(\ack)}
       }\bigl(\bar{\randvecy}^{(\ack)}\bigr)
       >\gamma\sub{f}
       },
       \prob{\frac{\mathrm{d}P^{(\nack)}}{\mathrm{d}P^{(\ack)}}
       \bigl(\bar{\randvecy}^{(\nack)}\bigr)
       \leq\gamma\sub{f}
       }
      \right)
    \end{IEEEeqnarray}
    \begin{IEEEeqnarray}{rCl}
      \bigcup_{\gamma\sub{f}\in \reals \cup \{\pm \infty\}}
      \left(
      \prob{
       \frac{\mathrm{d}P^{(\nack)}}{
       \mathrm{d}P^{(\ack)}
       }\bigl(\bar{\randvecy}^{(\ack)}\bigr)
       \geq\gamma\sub{f}
       },
       \prob{\frac{\mathrm{d}P^{(\nack)}}{\mathrm{d}P^{(\ack)}}\bigl(\bar{\randvecy}^{(\nack)}\bigr)<\gamma\sub{f}
       }
      \right).
    \end{IEEEeqnarray}
    Here, $\bar{\randvecy}^{(\nack)}\distas P^{(\nack)}$ and $\bar{\randvecy}^{(\ack)}\distas P^{(\ack)}$, where $P^{(\nack)}=P_{\bar{\randvecy}\given\bar{\randvecx}=\bar{f}(\nack)}$
    and $P^{(\ack)}=P_{\bar{\randvecy}\given\bar{\randvecx}=\bar{f}(\ack)}$.}
  \end{lem}

  \begin{IEEEproof}
    The result follows from a direct application of the Neyman-Pearson lemma~\cite{neyman33-01a}.
  \end{IEEEproof}

  Next, we present our achievability bound, which generalizes~\cite[Thm.~3]{polyanskiy11-08a} to the case of noisy feedback and of a finite number of transmission rounds.
  \begin{thm}\label{thm:HARQ_ach}
    Let $(\efa,\efb)$ be an achievable pair according to Lemma~\ref{lem:hypothesis-testing} for a given choice of $\nf$ and encoder for the feedback channel.
    Assume that $0\leq \efa+\efb \leq 1$.
     Fix three integers $M$, $\nmax$ and~$n$, and a real number $\gamma\sub{dec}$.
     Let $(\randvecx_1,\randvecx_2, \dots)$ be a \markblue{stationary memoryless} stochastic process where $\randvecx_\nu \in \mathcal{X}^n$ for every integer $\nu\geq 1$.
     \markblue{Let $P_{\randvecx}$ denote its marginal distribution, and assume that the mutual information $I(P_{\randvecx},P_{\randvecy\given\randvecx})$, where $P_{\randvecy\given\randvecx}$ is the channel law defined in~\eqref{eq:forward_channel}, is strictly positive.}

    \markblue{Also, let $\randvecy_{\nu}\distas P_{\randvecy\given\randvecx=\randvecx_{\nu}}$, $\nu\geq 1$, and consider a second stationary memoryless stochastic process $(\widetilde{\randvecx}_1,\widetilde{\randvecx}_2,\dots)$ with marginal distribution $P_{\randvecx}$ and independent of both $(\randvecx_1,\randvecx_2, \dots)$ and $(\randvecy_1,\randvecy_2, \dots)$.}
     %
     Finally define a sequence of information density functions $\mathcal{X}^{\nu n} \times \mathcal{Y}^{\nu n} \rightarrow \mathbb{R}$
     \begin{IEEEeqnarray}{rCl}\label{eq:info_dens}
       \imath_\nu\lro{\vecx^\nu,\vecy^\nu} &=& \log \frac{\mathrm{d}P_{\randvecy^\nu \given \randvecx^\nu}\lro{\vecy^\nu \given \vecx^\nu}}{\mathrm{d}P_{\randvecy^\nu}\lro{\vecy^\nu}}, \quad \nu=1,2,\dots
     \end{IEEEeqnarray}
  \end{thm}
  and two stopping times
  \begin{IEEEeqnarray}{rCl}
  {\tau} &=& \inf\{\nu\geq 1 :\imath_\nu\lro{\randvecx^\nu,\randvecy^\nu} \geq\gamma\sub{dec}\}\label{eq:tau},\\
  \widetilde{{\tau}} &=& \inf\{\nu\geq 1 :\imath_\nu\lro{\widetilde{\randvecx}^\nu,\randvecy^\nu} \geq\gamma\sub{dec}\}. \label{eq:taubar}
  \end{IEEEeqnarray}
  \ju{Then, there exists an~$\lro{\ell\sub{a}, M, \epsilon,  \nmax,n,\nf}$-\gls{vlsf} code whose  average service time $\ell\sub{a}$,  is upper-bounded by}
  \begin{equation}\label{eq:ell_ub}
    \ell\sub{a}\leq  \sum_{\nu=0}^{\nmax-1} \lro{ G_{\nu+1} - G_\nu} \prob{\tau > \nu}
  \end{equation}
  %
 and whose  average error probability is upper-bounded by
   \begin{IEEEeqnarray}{rCl}
      \epsilon &\leq& \sum_{\nu=1}^{\nmax}\xi_\nu \Bigl(\alpha_\nu \prob{ \tau > \nu} +  \lro{M-1}    \prob{\tau \geq \nu, \widetilde{\tau} = \nu} \Bigr) \label{eq:eps_ub}. \IEEEeqnarraynumspace
  \end{IEEEeqnarray}

  Here, $\alpha_\nu = \efb$ for $\nu = 1,\dots, \nmax-1$ and $\alpha_{\nmax} = 1$.
  Furthermore, $\xi_\nu = \lro{1-\efb}^{\nu-1}$ and
  \begin{IEEEeqnarray}{rCl}
  	G_\nu &=& \sum_{k=1}^{\nu-1} k \xi_k \efb {+} \xi_\nu \lrho{\sum_{k=\nu}^{\nmax-1}  k \efa^{k-\nu} \lro{1{-}\efa} {+} \nmax \efa^{\nmax-\nu}}\IEEEeqnarraynumspace \label{eq:h0-def}
  \end{IEEEeqnarray}
  for $\nu=1, \dots, \nmax$, whereas $G_0=0$.
  \begin{IEEEproof}
    See Appendix~\ref{app:thm_ach_harq}.
  \end{IEEEproof}

  Some remarks about our achievability bound are in order.
  As discussed in Appendix~\ref{app:thm_ach_harq}, our bound is based on a decoder that tracks the accumulated information density between each codeword and the received signal.
  The stopping rule is triggered whenever the accumulated information density exceeds the threshold $\gamma\sub{dec}$.
  The random variable $\tau$ in~\eqref{eq:tau} denotes the index of the first round in which the information density corresponding to the desired codeword exceeds the threshold, whereas $\widetilde{\tau}$ in~\eqref{eq:taubar} denotes the index of the first round in which a codeword different from the transmitted one exceeds the threshold.
  Clearly, the event $\tau> \widetilde{\tau}$  will correspond to an undetected error, provided that $\widetilde{\tau}\leq \nmax$ and no $\nack\to\ack$ error has occurred in the previous rounds.
  This is captured by the \markblue{second} term in the error-probability bound~\eqref{eq:eps_ub}.
  The \markblue{first} term in~\eqref{eq:eps_ub} captures instead the error resulting from a $\nack\to\ack$ event.

  Note that one recovers the bound reported in~\cite[Thm.~3]{polyanskiy11-08a} from the bound given in Theorem~\ref{thm:HARQ_ach} by setting $\efa=\efb=0$ and letting $\nmax\to\infty$.

  As shown in Appendix~\ref{app:exp_tau_rx}, the bound given in Theorem~\ref{thm:HARQ_ach} can be easily modified to account for the case in which the average latency at the receiver is the metric of interest, and $\ell\sub{a}$ gives a constraint on this quantity.
  One needs to replace~\eqref{eq:ell_ub} by
  \begin{equation}\label{eq:ell_ub_rx}
    \ell\sub{a}\leq 1 + \sum_{\nu=1}^{\nmax-1} \xi_{\nu} \prob{\tau > \nu}.
  \end{equation}

  \ju{In the \gls{urllc} literature, (see, e.g.,~\cite{Popovski18}), it is common to specify the latency $t$ of a packet transmission as
  \begin{IEEEeqnarray}{rCl}\label{eq:urllc_latency}
    t &=&
    \begin{cases}
      t_0' -t_0, &\text{if packet delivered error-free}\\
      \infty, &\text{otherwise}
    \end{cases}
  \end{IEEEeqnarray}
  where $t_0$ is the time instance the packet is made available to the transmitter and $t_0'$ is the time instance when the packet is delivered error-free by the receiver ($t_0'$ is not defined if the packet is not delivered).
  The \gls{urllc} service requirement can then be expressed as
  \begin{IEEEeqnarray}{rCl}
    \prob{t \leq t\sub{max}} &\geq & 1-\epsilon\sub{URLLC} \label{eq:urllc_error}
  \end{IEEEeqnarray}
  where $t\sub{max}$ is the latency requirement and $1-\epsilon\sub{URLLC}$ is the reliability requirement. }

\ju{
  The \gls{vlsf} scheme considered in this paper will satisfy the requirement~\eqref{eq:urllc_error} if $t\sub{max}\geq \nmax(n+\nf)$ channel uses and $\epsilon\sub{URLLC} \geq \epsilon$.
  However, in general, there is no simple relationship between $t$ as defined in~\eqref{eq:urllc_latency} and $\trx$, since $\trx$ is finite also when there are transmission errors.
}

  \section{Numerical Results}\label{sec:numerical_results}
  We show in this section how to use Theorem~\ref{thm:HARQ_ach} to \markblue{obtain guidelines on the design of} a \gls{harq}-based short-packet transmission system operating over a wireless channel.
  Specifically, we are interested in understanding the performance degradation due to noise in the feedback link.
  \markblue{Also, we seek prescriptions on how to choose the size of the codeword segments, the size of the repetition code that protects the feedback bit, the $\ack\rightarrow \nack$ and the $\nack\rightarrow \ack$ probability, and---for the fading case---the number of pilot symbols used to estimate the forward channel at the receiver.}

  Although our framework is general, we will consider for simplicity only the following two scenarios:
  \begin{enumerate*}
   \item both the forward and the feedback channel are real-valued \gls{biawgn} channels operating at possibly different SNR levels,
   \item both the forward and the feedback channel are Rayleigh block-memoryless fading channels operating at the same SNR level.
 \end{enumerate*}

 \subsection{The \gls{biawgn} scenario}\label{sec:biawgn}

  We assume that the additive noise has unit variance and that each transmit symbol belongs to the alphabet $\{-\sqrt{\rho},\sqrt{\rho}\}$, where $\rho$ denotes the SNR on the forward link.
  We also assume that the encoder for the feedback channel assigns the~$\nf$-dimensional vector $[\sqrt{\rho\sub{f}},\dots,\sqrt{\rho\sub{f}}]$ to the message~$\ack$ and the~$\nf$-dimensional vector $[-\sqrt{\rho\sub{f}},\dots,-\sqrt{\rho\sub{f}}]$ to $\nack$.
  Here, $\rho\sub{f}$ denotes the SNR on the feedback link.
  Under these assumptions, it follows from Lemma~\ref{lem:hypothesis-testing} that for a given Neyman-Pearson threshold~$\gamma\sub{f}$, the probabilities $\efa$ and $\efb$ can be expressed as
  \begin{IEEEeqnarray}{rCl}
    \efa &=& Q\lro{\sqrt{\nf \rho\sub{f}} + \gamma\sub{f}} \label{eq:np-awgn-1} \\
    \efb &=& Q\lro{\sqrt{\nf \rho\sub{f}} - \gamma\sub{f}} \label{eq:np-awgn-2}.
  \end{IEEEeqnarray}

  Next, we evaluate the bound in Theorem~\ref{thm:HARQ_ach} for a stationary memoryless input process with marginal distribution uniform over  $\{-\sqrt{\rho},\sqrt{\rho}\}$.
  For such a distribution, \eqref{eq:info_dens} reduces to
  \begin{IEEEeqnarray}{rCl}
    \imath_\nu\lro{\randvecx^\nu, \randvecy^\nu} &\distas& \sum_{i=1}^{\nu n} \log 2 - \log\lro{1+\exp\lro{-2 Z_i}} \label{eq:info_dens_biawgn}
  \end{IEEEeqnarray}
 where the $\lrbo{Z_i}$ are independent and $\mathcal{N}\lro{\rho, \rho}$ distributed.
  Since evaluating~\eqref{eq:eps_ub} directly is challenging, we use the following upper bound on the probability term $\prob{\tau \geq \nu, \widetilde{\tau} =\nu}$ in~\eqref{eq:eps_ub}:
  \begin{IEEEeqnarray}{rCl}
    \prob{\tau \geq \nu, \widetilde{\tau} =\nu} &\leq& \prob{\widetilde{\tau} =\nu}\\
    &=&\Ex{}{\exp\lro{- \imath_\nu\lro{\randvecx^\nu, \randvecy^\nu}}\mathbb{1}\{ \tau=\nu \}}.\label{eq:change_of_measure}\IEEEeqnarraynumspace
  \end{IEEEeqnarray}
 The equality in~\eqref{eq:change_of_measure}  follows from a change-of-measure argument (see~\cite[Eq.~(110)]{polyanskiy11-08a}).
 \markblue{Recall that the stopping time $\tau$, defined in~\eqref{eq:tau}, depends on the threshold $\gamma\sub{dec}$.}
 The resulting expression can be readily evaluated using Monte-Carlo methods.

 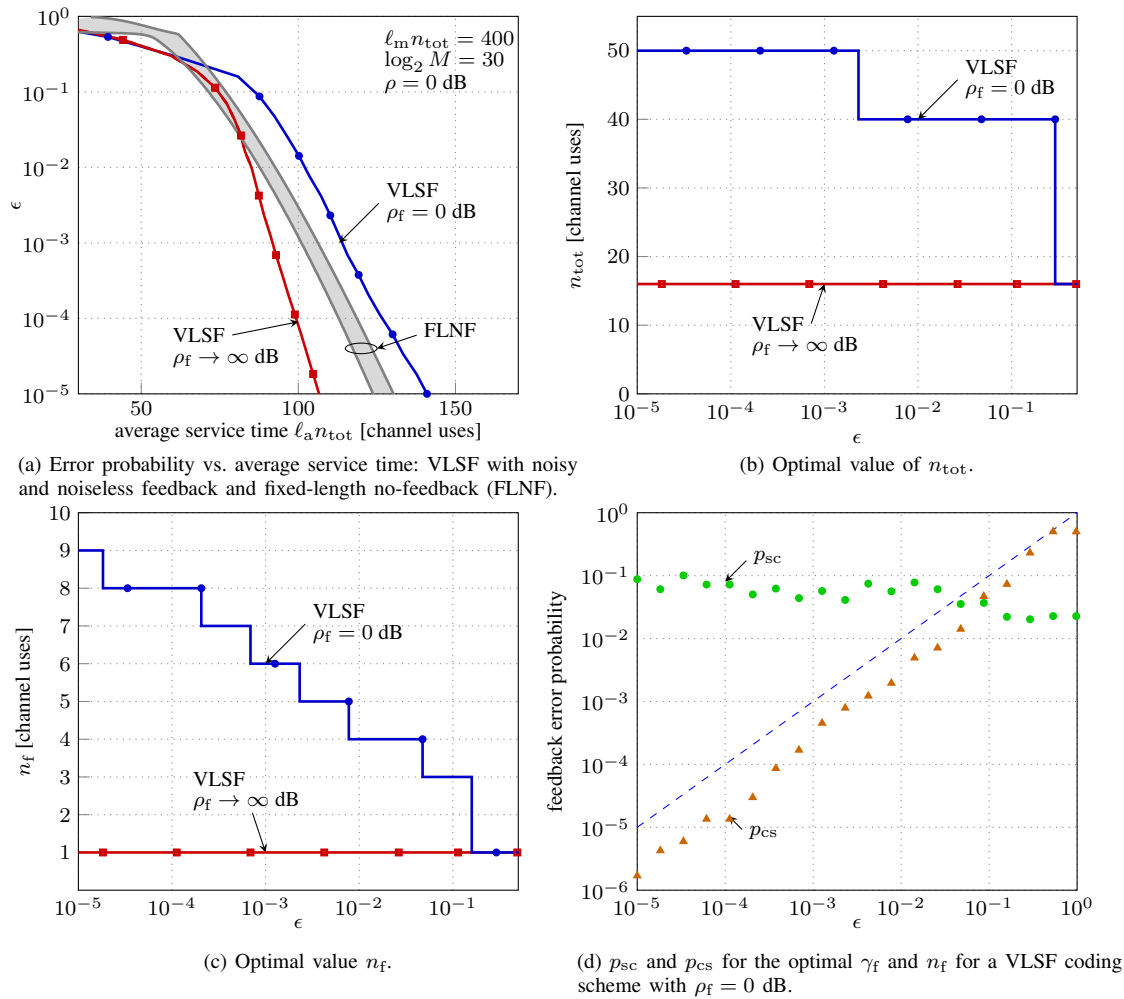
\begin{figure*}[t]
     \centering
         \begin{tikzpicture}
\pgfplotsset{
   scaled y ticks = false,
   height=\fwidth*0.4,
    width = 0.45*\textwidth,
    title style={yshift=-6pt,}
}
   \begin{groupplot}[
       group style={
       group name=my plots,
       group size = 2 by 2,
       vertical sep=45pt,
       horizontal sep=45pt,
       },
   ]
   \nextgroupplot[
         xlabel={average service time $\ell\sub{a}n\sub{tot}$ [channel uses]},
         ylabel = {$\epsilon$},
         grid=major,
         ytick style={draw=none},
         ymode = log,
         xmin = 30,
         xmax=170,
         ymin=1e-5,
         ymax = 1,
         legend style={at={(0.02,0.2)},anchor=west},
         ytick pos=left,
         xtick pos=bottom,
        ]

        \node[anchor=west] at (axis cs: 125,5e-1) (a) {$\nmax\ntot=400$};
        \node[below=0.3cm of a.west,anchor=west] (b) {$\log_2 M=30$};
        \node[below=0.3cm of b.west,anchor=west] (c) {$\rho=0$ dB};

   \addplot [mark size=1pt, mark=*,mark repeat=3,color=blue!80!black, line width=\linew] table [y index={1}, x index = {0}, col sep=comma] {./ell_vs_bler_VLNSF_envelope.csv};

   \addplot [color=red!80!black,mark size=1pt,mark=square*,mark repeat=3, line width=\linew] table [y index={1}, x index = {0}, col sep=comma] {./ell_vs_bler_VLSF_envelope.csv};

   \addplot [color=red!80!black,mark size=1pt,mark=square*,mark repeat=3, line width=\linew,dashed] table [y index={1}, x index = {0}, col sep=comma] {./VLSF_time_sharing_n_16_rho_0dB.csv};

   \addplot [name path = p1,mark = none, color=gray, line width=\linew] table [y index={1}, x index = {0}, col sep=comma] {./biawgn_fbl.csv};
   \addplot [name path = p2,mark = none, color=gray, line width=\linew] table [y index={2}, x index = {0}, col sep=comma] {./biawgn_fbl.csv};
   \addplot[gray,fill opacity=0.3] fill between[of=p2 and p1];

   \coordinate (pt1) at (axis cs: 113, 1e-3);
   \coordinate (pt2) at ($(pt1)+ (15pt,15pt)$);
   \draw[<-] (pt1)--(pt2) node[anchor=west] at ($(pt2) + (0pt,5pt)$) (aa) {VLSF};
   \node[below=0.3cm of aa.west,anchor=west] {$\rho\sub{f} = 0$ dB};

   \coordinate (pt1) at (axis cs: 100, 9e-5);
   \coordinate (pt2) at ($(pt1)+ (-25pt,-5pt)$);
   \draw[<-] (pt1)--(pt2) node[anchor=east] at ($(pt2) + (0pt,0pt)$) (aa) {VLSF};
   \node[below=0.3cm of aa.west,anchor=west] {$\rho\sub{f} \rightarrow \infty$ dB};

   \draw (axis cs:  120,4e-5)  ellipse  (6pt and 2pt);
   \coordinate (ut1) at (axis cs: 120,4e-5);
   \coordinate (pt1) at ($(ut1) + (6pt,0pt)$);
   \coordinate (pt2) at ($(pt1) + (15pt,5pt)$);
   \draw[<-] (pt1)--(pt2) node[anchor=west] at ($(pt2) + (0pt,2pt)$) {FLNF};

   \nextgroupplot[
         ylabel={$\ntot$ [channel uses]},
         xlabel = {$\epsilon$},
         grid=major,
         xmode = log,
         ymin = 0,
         xtick style={draw=none},
         ymax=55,
         xmin=1e-5,
         xmax = 5e-1,
         xtick pos=left,
         ytick pos=bottom,
         xlabel near ticks,
         ytick={0,10,20,30,40,50},
         yticklabels={$0$,$10$,$20$,$30$,$40$,$50$},
         xtick={1e-5,1e-4,1e-3,1e-2,1e-1},
         xticklabels={$10^{-5}$,$10^{-4}$,$10^{-3}$,$10^{-2}$,$10^{-1}$},
        ]

   \addplot [mark=square,mark size=1pt,mark repeat=3,const plot,color=red!80!black,line width=\linew] table [y index={2}, x index = {1}, col sep=comma] {./ell_vs_bler_VLSF_envelope.csv};

   \addplot [mark=*,mark size=1pt,mark repeat=3,const plot,color=blue!80!black,line width=\linew] table [y index={2}, x index = {1}, col sep=comma] {./ell_vs_bler_VLNSF_envelope.csv};

   \coordinate (pt1) at (axis cs: 1e-2,40);
   \coordinate (pt2) at ($(pt1)+ (15pt,15pt)$);
   \draw[<-] (pt1)--(pt2) node[anchor=west] at ($(pt2) + (0pt,5pt)$) (aa) {VLSF};
   \node[below=0.3cm of aa.west,anchor=west] {$\rho\sub{f} = 0$ dB};

   \coordinate (pt1) at (axis cs: 1e-3,16);
   \coordinate (pt2) at ($(pt1)+ (-5pt,-15pt)$);
   \draw[<-] (pt1)--(pt2) node[anchor=east] at ($(pt2) + (0pt,0pt)$) (aa) {VLSF};
   \node[below=0.3cm of aa.west,anchor=west] {$\rho\sub{f} \rightarrow \infty$ dB};

\nextgroupplot[
ylabel={$\nf$ [channel uses]},
xlabel={$\epsilon$},
grid=major,
xmode = log,
xtick style={draw=none},
ymin = 0,
ymax=10,
xmin=1e-5,
xmax = 5e-1,
xtick pos=left,
ylabel near ticks,
 ytick pos=bottom,
 ytick={1,2,3,4,5,6,7,8,9,10},
 yticklabels={$1$,$2$,$3$,$4$,$5$,$6$,$7$,$8$,$9$,$10$},
 xtick={1e-5,1e-4,1e-3,1e-2,1e-1},
 xticklabels={$10^{-5}$,$10^{-4}$,$10^{-3}$,$10^{-2}$,$10^{-1}$},
]
\addplot [mark=square,mark size=1pt,mark repeat=3,const plot,color=red!80!black,line width=\linew] table [y index={3}, x index = {1}, col sep=comma] {./ell_vs_bler_VLSF_envelope.csv};

\addplot [mark=*,mark size=1pt,mark repeat=3,const plot,color=blue!80!black,line width=\linew] table [y index={3}, x index = {1}, col sep=comma] {./ell_vs_bler_VLNSF_envelope.csv};

\coordinate (pt1) at (axis cs: 1e-3,6);
\coordinate (pt2) at ($(pt1)+ (15pt,15pt)$);
\draw[<-] (pt1)--(pt2) node[anchor=west] at ($(pt2) + (0pt,5pt)$) (aa) {VLSF};
\node[below=0.3cm of aa.west,anchor=west] {$\rho\sub{f} = 0$ dB};

\coordinate (pt1) at (axis cs: 1e-3,1);
\coordinate (pt2) at ($(pt1)+ (-5pt,15pt)$);
\draw[<-] (pt1)--(pt2) node[anchor=east] at ($(pt2) + (0pt,13pt)$) (aa) {VLSF};
\node[below=0.3cm of aa.west,anchor=west] {$\rho\sub{f} \rightarrow \infty$ dB};

   \nextgroupplot[
         xlabel={$\epsilon$ },
         xmode = log,
         ymode = log,
         ylabel = {feedback error probability},
         grid=major,
         xmin = 1e-5,
         xmax=1,
         ymin=1e-6,
         ymax = 1,
         ytick pos=left,
         xtick pos=bottom,
         ylabel near ticks,
         xtick style={draw=none},
         ytick style={draw=none},
        ]
  \addplot [mark=*,mark size=1pt,const plot,color=green!80!black,line width=\linew,only marks] table [y index={4}, x index = {1}, col sep=comma] {./ell_vs_bler_VLNSF_envelope.csv};

  \addplot [mark=triangle,mark size=1pt,const plot,color=orange!80!black,line width=\linew,only marks] table [y index={5}, x index = {1}, col sep=comma] {./ell_vs_bler_VLNSF_envelope.csv};

  \draw [blue,dashed] (1e-5,1e-5) -- (1,1);

  \coordinate (pt1) at (axis cs: 1.1e-4,1.5e-5);
  \coordinate (pt2) at ($(pt1)+ (5pt,-5pt)$);
  \draw[<-] (pt1)--(pt2) node[anchor=west] at ($(pt2) + (0pt,-1pt)$) (aa) {$\efb$};

  \coordinate (pt1) at (axis cs: 1e-4,8e-2);
  \coordinate (pt2) at ($(pt1)+ (8pt,8pt)$);
  \draw[<-] (pt1)--(pt2) node[anchor=west] at ($(pt2) + (0pt,1pt)$) (aa) {$\efa$};
\end{groupplot}

  \node[text width=0.45\textwidth,align=center,anchor=north,font=\footnotesize] at ([yshift=-5mm]my plots c1r1.south) {\subcaption{Error probability vs. average service time: VLSF with noisy and noiseless feedback and fixed-length no-feedback (FLNF).}\label{fig:biawgn1a}};
 \node[text width=0.45\textwidth,align=center,anchor=north,font=\footnotesize] at ([yshift=-5mm]my plots c2r1.south) {\subcaption{Optimal value of $\ntot$. }\label{fig:biawgn1b}};
 \node[text width=0.45\textwidth,align=center,anchor=north,font=\footnotesize] at ([yshift=-5mm]my plots c1r2.south) {\subcaption{Optimal value $\nf$.}\label{fig:biawgn1c}};
 \node[text width=0.45\textwidth,align=center,anchor=north,font=\footnotesize] at ([yshift=-5mm]my plots c2r2.south) {\subcaption{$\efa$ and $\efb$ for the optimal $\gamma\sub{f}$ and $\nf$ for a VLSF coding scheme with $\rho\sub{f}=0$ dB.}\label{fig:biawgn1d}};

\end{tikzpicture}%
     \caption{Optimal design of VLSF coding schemes for the bi-AWGN channel. }
     \label{fig:biawgn_1}
 \end{figure*}

 In the numerical simulations that \markblue{follow}, we require $\nmax$ and $\ntot$ to be integers and fix a target maximum latency $\nmax\ntot$, measured in channel uses, where $\ntot=n+\nf$.
 Then, for a given number of information bits $\log_2 M$, we use Theorem~\ref{thm:HARQ_ach} to obtain an upper bound on the error probability~$\epsilon$ achievable for a given \markblue{constraint}  $\ell\sub{a}\ntot$ on the average service time measured in channel uses.\footnote{\markblue{Specifically, Theorem~1 guarantees that whenever the average-service-time constraint (measured in number of rounds) is equal to the right-hand side of~\eqref{eq:ell_ub}, the error probability is smaller than the right-hand side of~\eqref{eq:eps_ub}.}}
 The bounds on the error probability reported in this section are optimized over the choice of the total number of symbols  per transmission round $n\sub{tot}$, the Neyman-Pearson threshold $\gamma\sub{f}$ in~\eqref{eq:np-awgn-1} and~\eqref{eq:np-awgn-2}, and the number of feedback symbols $\nf$, under the constraint that $n+\nf=n\sub{tot}$ and that $\nmax\ntot$ is equal to the targeted maximum latency.
 \markblue{The optimization is performed using a grid search algorithm.}
The \gls{vlsf} bounds reported in this section are obtained by time-sharing between the \gls{vlsf} scheme used to establish Theorem~\ref{thm:HARQ_ach} and a scheme in which the transmitter simply drops the packet, which results in~$\ell\sub{a}=0$ and~$\epsilon=1$.
Specifically, let $q\in\lrho{0,1}$ be the fraction of messages sent with the \gls{vlsf} scheme and, consequently, let $1-q$ be the fraction of messages that are dropped at the transmitter.
For each $\epsilon$ in Fig.~\ref{fig:biawgn_1}, $q$ is optimized to yield the smallest average service time.
Time sharing \markblue{turns out to be helpful in the high error-probability regime, i.e., when $\epsilon\geq 10^{-1}$.}


We start by considering the scenario in which the latency requirement is $\nmax\ntot=400$ channel uses, {$\rho=\rho\sub{f}=0$} dB, and $\log_2 M=30$ bits.
In Fig.~\ref{fig:biawgn1a}, we depict the upper bound on the error probability given in Theorem~\ref{thm:HARQ_ach} as a function of the \markblue{average-service-time constraint}.
For each error probability value, we present in Fig.~\ref{fig:biawgn1b} the optimum value of $\ntot$, in Fig.~\ref{fig:biawgn1c} the optimum value of~$\nf$, and in Fig.~\ref{fig:biawgn1d} the optimum value of $\efa$ and $\efb$.
For comparison, we also depict in Fig.~\ref{fig:biawgn1a} an upper bound on the error probability for the case in which the feedback link is noiseless, which is obtained by letting $\rho\sub{f}\to\infty$.
Note that, in this case, setting $\nf=1$ minimizes the error probability, as illustrated in Fig.~\ref{fig:biawgn1c}.
Finally, we plot an upper and a lower bound on the error probability achievable using a \gls{flnf} code, with blocklength $\ell\sub{a}\ntot$.
Specifically, the upper bound is  the random-coding union bound~\cite[Th. 16]{polyanskiy10-05a}, and the lower bound is the max-min bound~\cite[Th. 27]{polyanskiy10-05a}, evaluated using the saddlepoint approximation as described in~\cite{font-segura18-03}.

 Our results in Fig.~\ref{fig:biawgn1a} \markblue{illustrate the impact} of noise on the feedback channel on the error probability \markblue{(estimated on the basis of our upper bound)} \markblue{of the specific \gls{vlsf} coding scheme} considered in the paper.
 Consider for example a target error probability $\epsilon=10^{-5}$.
 When the feedback link is noiseless, the bound in Theorem~\ref{thm:HARQ_ach} yields a minimum average service time of~$106.6$ channel uses, an optimal value for~$\ntot$ of~$16$ channel uses, and an optimal value of~$\nf$ equal to~$1$.
 However, when noise in the feedback link is accounted for, the average service time increases to $141$ channel uses, the optimal value for $\ntot$ to $50$ channel uses, and the optimal value of $\nf$  to $9$ channel uses.
 The resulting average service time is larger than the one required by a \gls{flnf} coding scheme, which according to the achievability bound depicted in the figure, requires $130$ channel uses to operate at  $\epsilon=10^{-5}$.
 The performance degradation of the \gls{vlsf} coding scheme is caused by the resources that need to be allocated to the feedback link to decrease the frequency of $\nack\to\ack$ and $\ack\to\nack$ errors.
 Specifically, as shown in Fig.~\ref{fig:biawgn1d}, to achieve $\epsilon=10^{-5}$ it is sufficient to choose $\gamma\sub{f}=-1.65$, which results in $\efa=0.088$ and $\efb=1.7\times 10^{-6}$.
 Note that the  $\nack\to\ack$ event occurs with much smaller probability than the $\ack\to\nack$ event.

 Observe that the optimal number of channel uses $\ntot$ allocated on each round increases as the optimal number of feedback symbols $\nf$ increases.
 This has the positive effect of reducing the feedback signaling overhead; however, it has also the negative effect of reducing the maximum number of transmission rounds that are compatible with the given latency requirement.

\begin{figure}[t]
     \centering
         \begin{tikzpicture}
  \pgfplotsset{
      scaled y ticks = false,
      width=\textwidth*0.55,
      height=\textwidth*0.45,
       title style={yshift=-6pt,}
  }
 \begin{axis}
     [
      xlabel={feedback SNR $\rho\sub{f}$ [dB]},
      ylabel = {average service time $\ell\sub{a}\ntot$ [channel uses]},
      grid=none,
      xmin = 0,
      xmax=15,
      ymin=100,
      ymax = 150,
      ytick pos=left,
      xtick pos=bottom,
      ytick={100,110,120,130,140,150,160},
      yticklabels={$100$,$110$,$120$,$130$,$140$,$150$,$160$},
      grid = major
     ]
     \node[anchor=west] at (axis cs: 10,145) (b) {$\nmax \ntot=400$};
     \node[below=0.3cm of b.west,anchor=west] (c) {$\log_2 M=30$};
     \node[below=0.3cm of c.west,anchor=west] (d) {$\rho = 0$ dB}; 
     \node[below=0.3cm of d.west,anchor=west] (e) {$\epsilon = 10^{-5}$}; 

\addplot [mark size=1pt, mark=none,mark repeat=3,color=red!80!black, line width=1, forget plot] table [y index={1}, x index = {0}, col sep=comma] {./ell_vs_SNR.csv};

\addplot[color = green!80!black,  mark=*, mark size=1.5, only marks] table [y index={2}, x index = {0}, col sep=comma] {./ell_vs_SNR.csv};\label{plot_one}

\coordinate (pt1) at (axis cs: 4.2,106.6);
\coordinate (pt2) at ($(pt1)+ (8pt,-10pt)$);
\draw[<-] (pt1)--(pt2) node[anchor=north west] at ($(pt2) + (-2pt, 3pt)$) {VLSF, $\rho\sub{f}\rightarrow \infty$ dB};

\node[circle,inner sep=0pt] at (2.5,143) {$(\ntot, \nf) = (50,9)$};

\node[circle,inner sep=0pt] at (2,138) {$(50,7)$};
\node[circle,inner sep=0pt] at (3,133) {$(50,6)$};
\node[circle,inner sep=0pt] at (4,129) {$(50,5)$};
\node[circle,inner sep=0pt] at (5,126) {$(50,5)$};

\coordinate (pt1) at (axis cs: 5,120);
\coordinate (pt2) at (axis cs: 6,118);
\coordinate (pt3) at (axis cs: 7,117);
\coordinate (pt4) at ($(pt2)+ (10pt,15pt)$);
\draw[<-] (pt1)--(pt4) node[anchor=south] at ($(pt4) + (5pt,-2pt)$) {$(25,4)$};
\draw[<-] (pt2)--(pt4) ;
\draw[<-] (pt3)--(pt4) ;

\coordinate (pt1) at (axis cs: 8,113);
\coordinate (pt2) at (axis cs: 9,113);
\coordinate (pt3) at (axis cs: 10,113);
\coordinate (pt4) at (axis cs: 11,113);
\coordinate (pt5) at ($(pt2)+ (10pt,15pt)$);
\draw[<-] (pt1)--(pt5) node[anchor=south] at ($(pt5) + (5pt,-2pt)$) {$(25,3)$};
\draw[<-] (pt2)--(pt5) ;
\draw[<-] (pt3)--(pt5) ;
\draw[<-] (pt4)--(pt5) ;

\coordinate (pt1) at (axis cs: 12,108);
\coordinate (pt2) at (axis cs: 13,107);
\coordinate (pt3) at (axis cs: 14,107);
\coordinate (pt4) at (axis cs: 15,107);
\coordinate (pt5) at ($(pt2)+ (10pt,15pt)$);
\draw[<-] (pt1)--(pt5) node[anchor=south] at ($(pt5) + (5pt,-2pt)$) {$(16,1)$};
\draw[<-] (pt2)--(pt5) ;
\draw[<-] (pt3)--(pt5) ;
\draw[<-] (pt4)--(pt5) ;

\end{axis}

%

\end{tikzpicture}
     \caption{Average service time of a VLSF coding scheme as a function of the SNR $\rho\sub{f}$ on the feedback link.
               For reference, we illustrate the average service time for the noiseless case $\rho\sub{f}\to\infty$, for which setting $\nf=1$ and $n\sub{tot}=16$ is optimal. }
     \label{fig:biawgn_2}
 \end{figure}
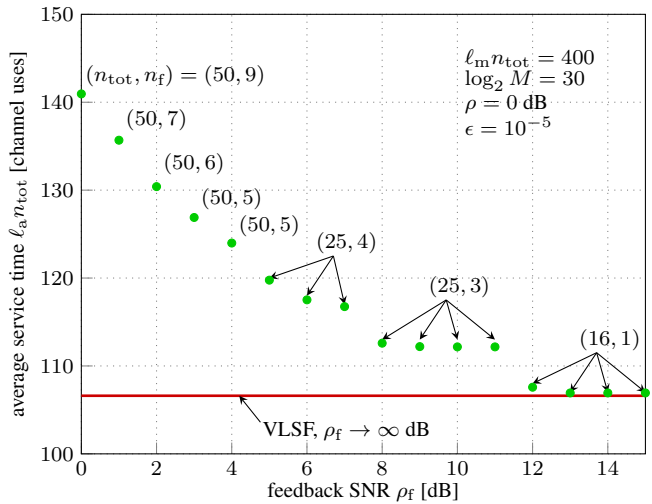

The performance of the \gls{vlsf} coding scheme for the case of noisy feedback can be improved by increasing the SNR $\rho\sub{f}$ on the feedback link.
This is illustrated in Fig.~\ref{fig:biawgn_2}, where we plot the average service time $\ell\sub{a}\ntot$ as function of the SNR $\rho\sub{f}$ on the feedback link.
As in Fig.~\ref{fig:biawgn_1}, we assume $\nmax\ntot=400$ channel uses, {$\rho=0$} dB, and $\log_2 M=30$ bits.
Furthermore, we focus on a target error probability $\epsilon=10^{-5}$.
The figure reveals that increasing the SNR $\rho\sub{f}$ to around $13$ dB yields an average service time close to that achievable in the noiseless-feedback case and optimal values of $n\sub{tot}$ and $\nf$ as in the noiseless-feedback case.

%
 %

 \subsection{The Rayleigh Fading Scenario}\label{sec:rayl_setup}
 We consider a setup in which the transmission in each round is through a quasi-static Rayleigh fading channel, i.e., the channel gain, which is Rayleigh distributed, stays constant over the transmission round.
 The fading coefficient is assumed to take independent realizations over different transmission rounds, \markblue{according to our block-memoryless assumption.}
 \markblue{Specifically, the input-output relation is given by}
 \begin{IEEEeqnarray}{rCl}
   \randvecy_{\nu} &=& H_{\nu} \randvecx_{\nu} + \randvecn_{\nu}.
 \end{IEEEeqnarray}
 Here, $\randvecx_{\nu}\in \complexset^n$ denotes the input and the output is $\randvecy_{\nu}\in\complexset^n$.
 The variable $H_{\nu}\sim\cgauss{0}{1}$ denotes the Rayleigh fading and $\randvecn_{\nu}\sim \cgauss{\veczero}{ \matI_{n}}$ denotes the AWGN.
 The random variables $\{H_{\nu} \}$ and $\{ \randvecn_{\nu} \}$ are assumed to be independent over $\nu$.\footnote{Independent fading realizations across transmission rounds can be achieved through, e.g., frequency hopping.}
 Furthermore, they do not depend on $\{\randvecx_\nu \}$.
 No \emph{a priori} knowledge of the realizations of $\{ H_{\nu} \}$ is assumed at \markblue{either} the transmitter or at the receiver.

We consider pilot-assisted transmission, which allows the receiver to acquire a noisy channel estimate.
 Specifically, similarly to~\cite{ostman18-12}, we consider inputs of the form $\randvecx_{\nu} = \lbrack\xp, \xd \rbrack$ where $\xp \in \complexset^{\np}$, $1\leq \np< n$ is a deterministic vector containing pilot symbols with $\vecnorm{\xp}^2 = \np \rho$, and $\xd \in \widetilde{\mathcal{X}}$ contains the $\nd = n-\np$ data symbols, drawn independently from a \gls{qpsk} constellation, i.e., $\widetilde{\mathcal{X}} = \{\sqrt{\rho}\exp\lro{\sqrt{-1} \frac{k\pi}{2}}, k=0,\dots, 3 \}^{\nd}$.
 \markblue{This choice is motivated by practical considerations.
 Better performance may be obtained using more sophisticated signaling schemes, for example based on the transmission of constant modulus vectors that are uniformly distributed on the power sphere, but at the price of higher receiver complexity.}

Let  $\vecy_\nu^{(\mathrm{p})}$ and $\vecy_\nu^{(\mathrm{d})}$ denote the received vectors corresponding to the pilot and the data symbols respectively.
Given $\vecx^{(\mathrm{p})}$ and $\vecy_\nu^{(\mathrm{p})}$, the receiver computes the \gls{ml} estimate of the fading realization as
\begin{IEEEeqnarray}{rCl}
  \widehat{h}_{\nu} &=& \frac{1}{\np \rho} \herm{(\xp)}\vecy_\nu^{(\mathrm{p})}.
\end{IEEEeqnarray}

We assume that the decoder treats the channel estimate as perfect  and computes for each codeword the following mismatched accumulated decoding metric
\begin{equation}\label{eq:generalized_info_dens}
  \jmath_\nu (\vecx^\nu,\vecy^\nu)=\sum_{k=1}^\nu \sum_{i=1}^{\nd} \log \frac{q(x_{k,i}^{(\mathrm{d})},y_{k,i}^{(\mathrm{d})})}{\Ex{}{q(X,y_{k,i}^{(\mathrm{d})})}}
\end{equation}
$\nu=1,\dots,\nmax$.\footnote{\markblue{The logarithmic term in~\eqref{eq:generalized_info_dens} is a special case of the generalized information density defined in~\cite[Eq.~(3)]{martinez11-02a}.
Indeed, this term can be obtained from~\cite[Eq.~(3)]{martinez11-02a} by setting $s=1$. We use the simpler expression provided in~\eqref{eq:generalized_info_dens} to avoid performing an optimization over $s$, which is time consuming.}}
Here, $x_{k,i}^{(\mathrm{d})}$ denotes the $i$th element of $\vecx_k^{(\mathrm{d})}$, and $q(x_{k,i}^{(\mathrm{d})},y_{k,i}^{(\mathrm{d})})$ is the \gls{snn} decoding metric
\begin{equation}\label{eq:scaled-nearest-neighbor}
  q(x_{k,i}^{(\mathrm{d})},y_{k,i}^{(\mathrm{d})})=\exp\lefto(-\abs{y_{k,i}^{(\mathrm{d})}-\widehat{h}_kx_{k,i}^{(\mathrm{d})}}^2\right)
\end{equation}
and $X$ in~\eqref{eq:generalized_info_dens} is uniformly distributed over $\widetilde{\mathcal{X}}$.
%
 Substituting~\eqref{eq:scaled-nearest-neighbor} into~\eqref{eq:generalized_info_dens}, we obtain
 \begin{IEEEeqnarray}{rCl}\label{eq:generalized_info_dens_final}
   \markblue{\jmath_\nu\lro{\vecx^\nu, \vecy^\nu} = \sum_{k=1}^\nu \sum_{i=1}^{\nd} -\abs{y^{(\mathrm{d})}_{k,i}-\widehat{h}_k x^{(\mathrm{d})}_{k,i}}^2 - \log \Ex{}{\exp( -\abs{y^{(\mathrm{d})}_{k,i}-\widehat{h}_k X}^2 )}.}
 \end{IEEEeqnarray}

To adapt Theorem~\ref{thm:HARQ_ach} to this mismatched-decoding setup, it is sufficient to replace $i_\nu$ in~\eqref{eq:tau} and~\eqref{eq:taubar} with \markblue{$\jmath_\nu$} in~\eqref{eq:generalized_info_dens_final}.
As in the \gls{biawgn} case, evaluating $\prob{\tau \geq \nu, \widetilde{\tau} =\nu}$ in~\eqref{eq:eps_ub} directly is challenging.
Hence, we resort to the following upper bound:
\begin{equation}\label{eq:simplification-bound-fading}
  \prob{\tau \geq \nu, \widetilde{\tau} =\nu} \leq \prob{\widetilde{\tau} =\nu} \leq \exp(-\gamma\sub{dec}).
\end{equation}
The proof of the last inequality can be found in Appendix~\ref{app:HARQ_ach_mm}.

 We model the feedback link in each transmission round as a quasi-static Rayleigh fading channel that is independent of the forward channel.
 The input-output relation in round $\nu=1,2,\dots, \nmax$ is given as
 \begin{IEEEeqnarray}{rCl}
   \bar{\randvecy}_\nu &=& \bar{H}_\nu \bar{\randvecx}_\nu + \bar{\randvecn}_\nu
 \end{IEEEeqnarray}
 where $\bar{\randvecx}_\nu \in \bar{\mathcal{X}}^{\nf}$ denotes the input to the feedback channel in round $\nu$ and $\bar{\randvecy}_{\nu}\in \bar{\mathcal{Y}}^{\nf}$ denotes the corresponding output.
 As before, $\bar{H}_\nu \sim \cgauss{0}{1}$ denotes the Rayleigh fading and $\bar{\randvecn}_\nu\sim \cgauss{\veczero}{\matI_{\nf}}$ denotes the \gls{awgn}.
 Again, $\{ \bar{H}_\nu \}$ and $\{ \bar{\randvecn}_\nu \}$ are mutually independent and also independent over~$\nu$, and do not depend on $\{\bar{\randvecx}_\nu\}$.
 Furthermore, no \emph{a priori} knowledge of the realizations of $\{ \bar{H}_\nu\}$ is assumed at the transmitter and at the receiver.
 Throughout this section, we assume that the forward and the feedback channels operate at the same SNR $\rho$.

 Since the channel is not known, we assume that the receiver uses on-off keying to signal the feedback bit.
 Specifically, the $\nack$ and an $\ack$ messages are mapped to the $\nf$-dimensional vectors $\lrho{0, \dots, 0}$ and $[ \sqrt{\rho}, \dots, \sqrt{\rho} ]$, respectively.
 \markblue{Also in this case, our choice is motivated by practical considerations; better-performing signaling schemes may be devised at the cost of higher complexity.}
 To perform binary-hypothesis testing based on the received vector $\bar{\vecy}_{\nu}$, the transmitter uses the noncoherent metric\footnote{In what follows, we omit the index $\nu$ to keep notation compact.}
 \begin{IEEEeqnarray}{rCl}
   \log \frac{P_{\bar{\randvecy} \given \bar{\randvecx} }\lro{\bar{\vecy} \given \bar{f}(\mathrm{c})}}{P_{\bar{\randvecy} \given \bar{\randvecx}}\lro{\bar{\vecy} \given \bar{f}(\mathrm{s}) }} &=& \log(1+\rho\nf)-\frac{\rho}{1+\nf \rho } \Bigl|{ \sum_{i=1}^{\nf} \bar{y}_i}\Bigr|^2.
 \end{IEEEeqnarray}
 Hence, we have that
 \begin{IEEEeqnarray}{rCl}
 \efa &=& P_{\bar{\randvecy} \given \bar{\randvecx} = \bar{f}(\ack)}\lrho{ \Bigl|\sum_{i=1}^{\nf} \bar{Y}_i\Bigr|^2 \leq \gamma\sub{f} } \\
 &=&
 1- \exp\lro{-\frac{\gamma\sub{f}}{\nf \lro{\nf \rho +1}}}
 \end{IEEEeqnarray}
 and
 \begin{IEEEeqnarray}{rCl}
 \efb &=& P_{\bar{\randvecy} \given \bar{\randvecx} = \bar{f}(\nack)}\lrho{ \Bigl|\sum_{i=1}^{\nf} \bar{Y}_i\Bigr|^2 > \gamma\sub{f} }\\
 &=&
 \exp\lro{-\gamma\sub{f}/\nf}.
 \end{IEEEeqnarray}


 \begin{figure}[t]
     \centering
         \begin{tikzpicture}
\pgfplotsset{
   scaled y ticks = false,
   height=\fwidth*0.4,
    width = 0.45*\textwidth,
    title style={yshift=-6pt,}
}
   \begin{groupplot}[
       group style={
       group name=my plots,
       group size = 2 by 2,
       vertical sep=45pt,
       horizontal sep=45pt,
       },
   ]
   \nextgroupplot[
    xlabel={average service time $\ell\sub{a} \ntot$ [channel uses]},
    ylabel = {$\epsilon$},
    grid=major,
    xmin = 25,
    xmax=150,
    ymin=1e-5,
    ymax = 1,
    ymode=log,
    ytick pos=left,
    xtick pos=bottom,
    ytick style={draw=none},
   ]
   \node[anchor=east] at (axis cs: 135,6e-1) (a) {$\nmax\ntot=400$};
   \node[below=0.3cm of a.west,anchor=west] (b) {$\log_2 M=30$};
   \node[below=0.3cm of b.west,anchor=west] (c) {$\rho=10$ dB};

\addplot [mark size=1pt, mark=*,mark repeat=3,color=blue!80!black, line width=\linew, forget plot] table [y index={1}, x index = {0}, col sep=comma] {./Rayl_ell_vs_bler_VLNSF_envelope.csv};


\addplot[color = red!80!black,  mark=square, mark size=1pt, mark repeat=3,line width=\linew] table [y index={1}, x index = {0}, col sep=comma] {./Rayl_ell_vs_bler_VLSF_envelope.csv};

\addplot[color = gray!80!black,  mark=square, mark size=1pt, mark repeat=1,line width=\linew] table [y index={1}, x index = {0}, col sep=comma] {./FLNF_Rayleigh.csv};

\coordinate (pt1) at (axis cs: 71,5e-3);
\coordinate (pt2) at ($(pt1)+ (25pt,10pt)$);
\draw[<-] (pt1)--(pt2) node[anchor=west] at ($(pt2) + (0pt,0pt)$) (aa) {VLSF};
\node[below=0.3cm of aa.west,anchor=west] {$\rho\sub{f} = 10$ dB};

\coordinate (pt1) at (axis cs: 58, 1e-2);
\coordinate (pt2) at ($(pt1)+ (-20pt,-10pt)$);
\draw[<-] (pt1)--(pt2) node[anchor=east] at ($(pt2) + (0pt,0pt)$) (aa) {VLSF};
\node[below=0.3cm of aa.west,anchor=west] {$\rho\sub{f} \rightarrow \infty$ dB};

\coordinate (pt1) at (axis cs: 100, 9e-4);
\coordinate (pt2) at ($(pt1)+ (15pt,5pt)$);
\draw[<-] (pt1)--(pt2) node[anchor=west] at ($(pt2) + (0pt,0pt)$) (aa) {FLNF};

   \nextgroupplot[
         ylabel={$\ntot$ [channel uses]},
         xlabel = {$\epsilon$},
         grid=major,
         xmode = log,
         ymin = 0,
         xtick style={draw=none},
         ymax=55,
         xmin=1e-5,
         xmax = 2e-1,
         xtick pos=left,
         ytick pos=bottom,
         xlabel near ticks,
         ytick={0,10,20,30,40,50},
         yticklabels={$0$,$10$,$20$,$30$,$40$,$50$},
        ]

   \addplot [mark=square,mark size=1pt,mark repeat=3,const plot,color=red!80!black,line width=\linew] table [y index={2}, x index = {1}, col sep=comma] {./Rayl_ell_vs_bler_VLSF_envelope.csv};

   \addplot [mark=*,mark size=1pt,mark repeat=3,const plot,color=blue!80!black,line width=\linew] table [y index={2}, x index = {1}, col sep=comma] {./Rayl_ell_vs_bler_VLNSF_envelope.csv};

   \addplot [mark=*,mark size=1pt,mark repeat=1,const plot,color=gray!80!black,line width=\linew] table [y index={3}, x index = {1}, col sep=comma] {./FLNF_Rayleigh.csv};

   \coordinate (pt1) at (axis cs: 2e-4,45);
   \coordinate (pt2) at ($(pt1)+ (15pt,5pt)$);
   \draw[<-] (pt1)--(pt2) node[anchor=west] at ($(pt2) + (2pt,3pt)$) (aa) {VLSF};
   \node[below=0.3cm of aa.west,anchor=west] {$\rho\sub{f} = 10$ dB};

   \coordinate (pt1) at (axis cs: 5e-5,40);
   \coordinate (pt2) at ($(pt1)+ (15pt,-25pt)$);
   \draw[<-] (pt1)--(pt2) node[anchor=north] at ($(pt2) + (0pt,0pt)$) (aa) {VLSF};
   \node[below=0.3cm of aa.west,anchor=west] {$\rho\sub{f} \rightarrow \infty$ dB};

   \coordinate (pt1) at (axis cs: 1e-3,16);
   \coordinate (pt2) at ($(pt1)+ (15pt,-15pt)$);
   \draw[<-] (pt1)--(pt2) node[anchor=north] at ($(pt2) + (0pt,0pt)$) (aa) {FLNF};

\nextgroupplot[
ylabel={channel uses},
xlabel={$\epsilon$},
grid=major,
xmode = log,
xtick style={draw=none},
ymin = 0,
ymax=7,
xmin=1e-5,
xmax = 1,
xtick pos=left,
ylabel near ticks,
 ytick pos=bottom,
 ytick={1,2,3,4,5,6,7,8,9,10},
 yticklabels={$1$,$2$,$3$,$4$,$5$,$6$,$7$,$8$,$9$,$10$},
]
\addplot [mark=square,mark size=1pt,mark repeat=3,const plot,color=red!80!black,line width=\linew] table [y index={3}, x index = {1}, col sep=comma] {./Rayl_ell_vs_bler_VLSF_envelope.csv};

\addplot [mark=square,mark size=1pt,mark repeat=3,const plot,color=red!80!black,line width=\linew, dashed] table [y index={4}, x index = {1}, col sep=comma] {./Rayl_ell_vs_bler_VLSF_envelope.csv};

\addplot [mark=*,mark size=1pt,mark repeat=3,const plot,color=blue!80!black,line width=\linew] table [y index={3}, x index = {1}, col sep=comma] {./Rayl_ell_vs_bler_VLNSF_envelope.csv};

\addplot [mark=*,mark size=1pt,mark repeat=3,const plot,color=blue!80!black,line width=\linew,dashed] table [y index={4}, x index = {1}, col sep=comma] {./Rayl_ell_vs_bler_VLNSF_envelope.csv};

\coordinate (pt1) at (axis cs: 1.2e-1,5.5);
\draw (pt1) ellipse  (8pt and 3pt);
\coordinate (pt2) at ($(pt1)+ (5pt,2pt)$);
\coordinate (pt3) at ($(pt2)+ (10pt,10pt)$);
\draw[<-] (pt2)--(pt3) node at ($(pt3) + (5pt,0pt)$) {$\np$};

\coordinate (pt1) at (axis cs: 9e-2,1.5);
\draw (pt1) ellipse  (3pt and 14pt);
\coordinate (pt2) at ($(pt1)+ (1pt,-12pt)$);
\coordinate (pt3) at ($(pt2)+ (5pt,-5pt)$);
\draw[<-] (pt2)--(pt3) node at ($(pt3) + (5pt,-1pt)$) {$\nf$};

\coordinate (pt1) at (axis cs: 4.1e-4,4);
\coordinate (pt2) at ($(pt1)+ (10pt,15pt)$);
\draw[<-] (pt1)--(pt2) node[anchor=east] at ($(pt2) + (25pt,10pt)$) (aa) {VLSF};
\node[below=0.3cm of aa.west,anchor=west] {$\rho\sub{f} = 10$ dB};
\coordinate (pt3) at (axis cs: 2e-4,6.5);
\coordinate (pt4) at ($(pt2)+ (0pt,10pt)$);
\draw[<-] (pt3)--(pt4);

\coordinate (pt1) at (axis cs: 1e-3,1);
\coordinate (pt2) at ($(pt1)+ (-10pt,15pt)$);
\draw[<-] (pt1)--(pt2) node[anchor=east] at ($(pt2) + (0pt,13pt)$) (aa) {VLSF};
\node[below=0.3cm of aa.west,anchor=west] {$\rho\sub{f} \rightarrow \infty$ dB};
\coordinate (pt3) at (axis cs: 1e-1,5);
\coordinate (pt4) at ($(pt2)+ (0pt,10pt)$);
\draw[<-] (pt3)--(pt4);

   \nextgroupplot[
         xlabel={$\epsilon$ },
         xmode = log,
         ymode = log,
         ylabel = {feedback error probability},
         grid=major,
         xmin = 1e-5,
         xmax=1,
         ymin=1e-6,
         ymax = 1,
         ytick pos=left,
         xtick pos=bottom,
         ylabel near ticks,
         xtick style={draw=none},
         ytick style={draw=none},
        ]
  \addplot [mark=*,mark size=1pt,const plot,color=green!80!black,line width=\linew,only marks] table [y index={5}, x index = {1}, col sep=comma] {./Rayl_ell_vs_bler_VLNSF_envelope.csv};

  \addplot [mark=triangle,mark size=1pt,only marks,color=orange!80!black,line width=\linew] table [y index={6}, x index = {1}, col sep=comma] {./Rayl_ell_vs_bler_VLNSF_envelope.csv};

  \draw [blue,dashed] (1e-5,1e-5) -- (1,1);

  \coordinate (pt1) at (axis cs: 4.5e-3,1.3e-1);
  \coordinate (pt2) at ($(pt1)+ (5pt,-10pt)$);
  \draw[<-] (pt1)--(pt2) node[anchor=north] at ($(pt2) + (3pt,1pt)$) (aa) {$\efa$};

  \coordinate (pt1) at (axis cs: 4.6e-3,2.2e-3);
  \coordinate (pt2) at ($(pt1)+ (8pt,-14pt)$);
  \draw[<-] (pt1)--(pt2) node[anchor=north] at ($(pt2) + (3pt,1pt)$) (aa) {$\efb$};
\end{groupplot}

  \node[text width=0.45\textwidth,align=center,anchor=north,font=\footnotesize] at ([yshift=-5mm]my plots c1r1.south) {\subcaption{Error probability vs. average service time: VLSF with noisy and noiseless feedback and FLNF.}\label{fig:ray1a}};
 \node[text width=0.45\textwidth,align=center,anchor=north,font=\footnotesize] at ([yshift=-5mm]my plots c2r1.south) {\subcaption{Optimal value of $\ntot$. }\label{fig:ray1b}};
 \node[text width=0.45\textwidth,align=center,anchor=north,font=\footnotesize] at ([yshift=-5mm]my plots c1r2.south) {\subcaption{Optimal values of $\nf$ and $n\sub{p}$.}\label{fig:ray1c}};
 \node[text width=0.45\textwidth,align=center,anchor=north,font=\footnotesize] at ([yshift=-5mm]my plots c2r2.south) {\subcaption{$\efa$ and $\efb$ for the optimal $\gamma\sub{f}$ and $\nf$ for  a VLSF coding scheme with $\rho\sub{f}=10$ dB.}\label{fig:ray1d}};

\end{tikzpicture}%
     \caption{Optimal design of \gls{vlsf} coding schemes for the block-memoryless Rayleigh fading channel.}
     \label{fig:rayl_1}
 \end{figure}
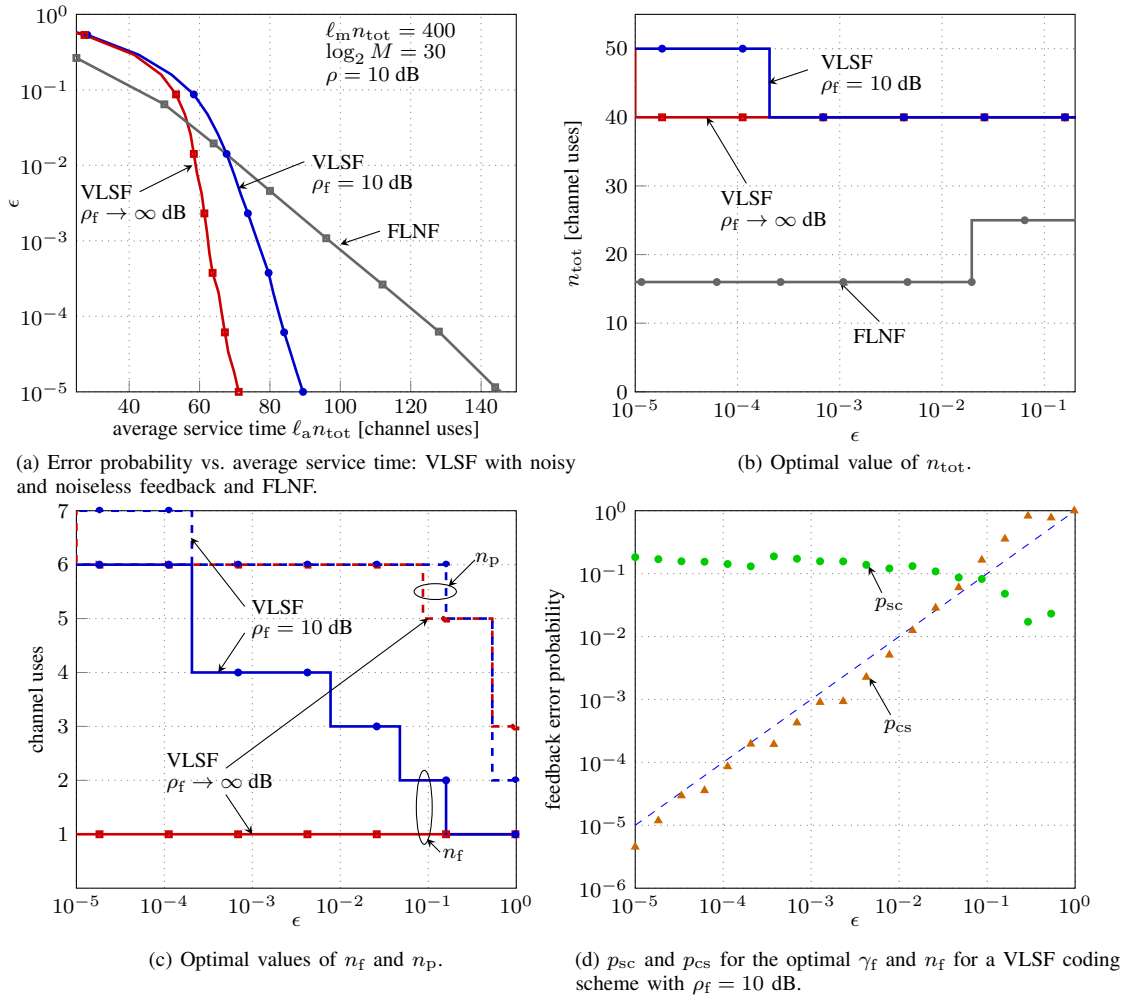

We consider again the scenario in which $\nmax\ntot=400$ channel uses and $\log_2 M=30$.
The SNR in both the forward and the feedback links is set to $10\dB$.
We optimize the bound on the error probability over $\ntot$, $\nf$, $\gamma\sub{f}$, and also over the number of pilot symbols $\np$.
Furthermore, as in the \gls{biawgn} case, we utilize a time-sharing strategy to tighten the achievability bound for high error probabilities.
We depict in Fig.~\ref{fig:ray1a} the error probability of the VLSF coding scheme as a function of the average service time.
For comparison, we also illustrate the error probability for the case of noiseless feedback, and an achievability bound on the error probability for the \gls{flnf} case based on~\cite[Th.1]{martinez11-02a}.
To obtain the \gls{flnf} curve, we assume that, for a fixed $\ntot$, a fixed-length scheme is used over $\ell\sub{a}$ consecutive coherence intervals, with $\ell\sub{a}$ being an integer.
The final curve is obtained by optimizing over $\ntot$.

We see in Fig.~\ref{fig:ray1a} that the presence of noise in the feedback link causes again a significant degradation of the \markblue{error probability, estimated on the basis of our upper bound.}
For example, for the case of noiseless feedback, the minimum average service time required to achieve $\epsilon=10^{-5}$ is $71$ channel uses when $\ntot=50$  (see Fig.~\ref{fig:ray1b}).
The average service time increases to $89.4$ channel uses, achieved again for $\ntot=50$, when noise in the feedback link is taken into account.
The minimum number of channel uses required by an \gls{flnf} scheme is $144$ channel uses, which is achieved for $\ntot=16$.
Differently from the bi-AWGN case,  this is significantly larger than the one achievable with the \gls{vlsf} scheme, even when noise in the feedback link is taken into account.

The reason behind the superior performance of the \gls{vlsf} scheme in the fading case is its implicit rate-adaptation capability~\cite{wu10-04}.
Specifically, in the \gls{flnf} case, one needs to choose the number $\ell\sub{a}$ of coherence intervals to code over in a conservative way, to mitigate the impact of deep fades.
On the contrary, in the \gls{vlsf} setup, this choice is made adaptively on the basis of the instantaneous fading realizations.

In Fig.~\ref{fig:ray1c}, we illustrate the optimal number of pilot symbols and feedback symbols.
We see that the number of pilot symbols increases as the target error probability decreases.
This is due to the additional resources required to convey~$\mathrm{s}$ and~$\mathrm{c}$ reliably, which weakens the \gls{vlsf} code on the forward link and, hence, increases the need of an accurate channel estimation.

In Fig.~\ref{fig:ray1d}, we plot~$\efa$ and~$\efb$ for the optimal choice of~$\gamma\sub{f}$ and~$\nf$.
Observe that, for $\epsilon > 5\times 10^{-2}$, the probability $\efb$ is actually greater than the target error probability $\epsilon$.
This is because, when $\epsilon > 5\times 10^{-2}$, a single transmission round suffices and a $\mathrm{c}$ symbol is never transmitted on the feedback channel.
As the target error probability $\epsilon$ decreases, retransmissions become necessary and, consequently, $\efb$ becomes smaller than $\epsilon$.

 \section{Conclusion}\label{sec:conclusion}
 We have generalized the achievability bound for \gls{vlsf} coding schemes presented in~\cite[Thm.~3]{polyanskiy11-08a} to the case in which the feedback channel is noisy and the feedback delay is accounted for.
\markblue{
Numerical results based on the bound provided in Theorem~\ref{thm:HARQ_ach} suggest that the estimate on the minimum average service time obtainable by using~\cite[Thm.~3]{polyanskiy11-08a} is inaccurate when noise in the feedback link is accounted for.}
 For example, in the \gls{biawgn} case, when the SNR is $0$ dB, the maximum latency is $400$ channel uses, and the target packet error probability is $10^{-5}$, Theorem~\ref{thm:HARQ_ach} \markblue{yields an estimate} of the average service time achievable with \gls{vlsf} coding schemes that  is larger than that achievable with \gls{flnf} coding schemes, once noise in the feedback link is accounted for (see Fig.~\ref{fig:biawgn_1}).
 In the fading case, however, under the same latency and reliability requirements, \markblue{Theorem~\ref{thm:HARQ_ach} suggests that} \gls{vlsf} coding schemes are preferable to \gls{flnf} schemes even when the feedback link is noisy (see Fig.~\ref{fig:rayl_1}).
\markblue{The intuition is that} \gls{vlsf} schemes utilize the available diversity more efficiently from a service-time perspective.

\markblue{Hence}, our analysis \markblue{suggests} that care must be exercised in utilizing simplifying assumptions such as perfect acknowledgment reception in the design of \gls{urllc} systems.

 As illustrated in Fig.~\ref{fig:biawgn1c} and Fig.~\ref{fig:biawgn1d}, to compensate for noise in the feedback link, which makes (uncoded) acknowledgments unreliable, \markblue{Theorem~\ref{thm:HARQ_ach} suggests that} one has to allocate additional resources to the feedback channel.
 This implies that fewer resources are available on the forward channel, which yields an overall performance degradation for small values of the average service time.
 As shown in Fig.~\ref{fig:biawgn_2}, one can compensate for such losses by transmitting the acknowledgments at a higher power level.
 This, however, may be unfeasible in bidirectional nonsporadic communications, where the acknowledgments are typically \markblue{piggybacked on packets} transmitted on the reverse data link.

\markblue{We hasten to add that our observations are entirely based on an upper bound on the error probability achievable using VLSF codes, whose tightness we are not able to assess.
Indeed, obtaining a tight converse bound for the case of noisy stop feedback is an open problem.
In fact, even for the case of a noiseless feedback link, no \gls{vlsf} converse result is known to the authors beyond the one obtainable by assuming full feedback.
This implies in particular that the tightness of~\cite[Thm.~3]{polyanskiy11-08a}---which is the bound we generalized in this paper---is also difficult to assess.}

{For the case of a noiseless feedback link with $\nmax=\infty$ and $n=1$, Theorem~\ref{thm:HARQ_ach} is known to be tight up to second order as the average blocklength grows large~\cite[Th. 2]{polyanskiy11-08a}.}
 Investigating whether a similar result can be established for the noisy-feedback case is left for future work.

 \markblue{Our analysis is based on the simplifying assumption that the decoder is perfectly aware of whether each codeword segment contains a new information message or just incremental redundancy.
 One way to relax this assumption is to protect the binary flag conveying this information using a repetition code.
 Then, the probability that the transmitter and the receiver fall out of synchronization can be computed using Lemma~\ref{lem:hypothesis-testing}.
An extension of Theorem~\ref{thm:HARQ_ach} to account for such an error event is nontrivial and is left for future work.
 }

 \appendices
 \section{Proof of Theorem \ref{thm:HARQ_ach}}
 \label{app:thm_ach_harq}
 %
 \markblue{Similar to~\cite[Thm.~3]{polyanskiy11-08a}, we start by defining} a random variable $U$ on the set~\footnote{Similar to~\cite[Section~II]{Trillinsgaard18-12} (see also~\cite[Thm.~19]{polyanskiy11-08a}), one can reduce the cardinality of this random variable to $2$.}
 \begin{equation}\label{eq:prob_space_auxiliary_rv}
   \setU=\underbrace{\setX^{\infty} \times \cdots \times \setX^{\infty}}_{M \text{ times}}
 \end{equation}
 with probability mass function
 \begin{equation}
     P_U=\underbrace{P_{\randvecx^\infty} \times \cdots \times P_{\randvecx^{\infty}}}_{M \text{ times}}
 \end{equation}
 where $P_{\randvecx^\infty}$ denotes the distribution of the \markblue{stationary memoryless stochastic} process $\{\randvecx_1,\randvecx_2,\dots\}$.
 Each realization of $U$ produces $M$ infinite-dimensional codewords $[\randvecc_1(w),\randvecc_2(w),\dots]$, $w=1,\dots,M$ where  each codeword segment  $\randvecc_{\nu}(w)$ belongs to $\setX^n$, $\nu=1,2,\dots$.
 The encoder $f_\nu$ maps the message $w$ to the codeword segment $\randvecc_\nu(w)$.

\markblue{
 We shall next follow the so-called random coding approach and characterize the average error probability and the average service time, averaged over all codebooks constructed according to this procedure.
 Note that, contrary to the common application of the random coding approach, establishing an upper bound on the average service time and the average error probability averaged over all codebooks does not imply the existence of a single codebook in the ensemble that satisfies both constraints.
 This problem is solved by the introduction of the random variable $U$, which enables the use of randomized coding strategies: each time a new message is transmitted, a new codebook is drawn from the ensemble.
 As shown in~\cite[Section~II]{Trillinsgaard18-12} (see also~\cite[Thm.~19]{polyanskiy11-08a}), it turns out sufficient to perform randomization across two codebooks.
 This implies that the cardinality of the set over which $U$ is defined can be reduced to $2$.
 In practice, one could implement randomization across the two codebooks by equipping the transmitter and the receiver with a pseudo number generator, and by ensuring that the generators are initialized using the same seed.
}

\markblue{We now continue with the proof.}
 As detailed in Algorithm~\ref{alg:algorithm}, the transmitter is also equipped with a stopping rule, which defines a stopping time $\ttx$ as follows:
 \begin{IEEEeqnarray}{rCl}
   \label{eq:tx_tau}
 	\ttx = \min\{ \nmax, \min\{\nu : \widehat{F}_\nu = \mathrm{s} \}\}.
 \end{IEEEeqnarray}
 Here, we use the convention that the minimum of an empty set is $\infty$.

 At the decoding side, we consider the following stopping rule: stop at round $\nu$ if $\imath_\nu\lro{\randvecc^\nu(w),\randvecy^\nu} \geq\gamma\sub{dec}$ for some $w$.
 Let now

 \begin{IEEEeqnarray}{rCl}
 	\tau_w \!=\! \min\lrbo{ \nu: \imath_\nu\lro{\randvecc^\nu(w), \randvecy^\nu} \geq \gamma\sub{dec}}
 \end{IEEEeqnarray}

 and let
 \begin{IEEEeqnarray}{rCl}
   \tau\sub{dec} &=& \min\lrbo{\tau_1, \dots, \tau_M}.\label{eq:tau_star}
 \end{IEEEeqnarray}
 Finally, let\footnote{Recall that the \markblue{decoder is assumed to know the feedback bit estimate at the transmitter.}}
 \begin{IEEEeqnarray}{rCl}\label{eq:tau_rx_def}
   \trx &=& \min\{\tau\sub{dec},\ttx +1, \nmax\}
 \end{IEEEeqnarray}
 be the stopping time at the decoder.
 If $\trx=\tau\sub{dec}$, the decoder sets $\widehat{W}=\max\{w \sothat \tau_{w}=\tau\sub{dec}\}$.
 Otherwise it sets $\widehat{W}=\mathrm{e}$.
 In words, an erasure is declared if no codeword results in a threshold crossing or if a $\nack\to\ack$ error occurs.
 Otherwise, the index of the codeword that resulted in a threshold crossing is taken as the message estimate.
 If a threshold crossing occurs for two or more codewords,  the codeword with the largest index is chosen.
 \markblue{Note that differently from~\cite[Thm.~3]{polyanskiy11-08a}, where one is interested in characterizing the expected value of $\tau\sub{dec}$, in our setup the quantity of interest is the expected value of $\tau\sub{tx}$, whose dependence on $\tau\sub{dec}$ will be made explicit next.}

 \markblue{Assume that the transmitted codeword has index $w'$.
 Since both the input process and the channel law are stationary and block memoryless, the accumulated information density $\imath_\nu\lro{\randvecc^\nu(w'), \randvecy^\nu}$, $\nu=1,2,\dots$ describes a random walk.
 Furthermore, since $\Ex{}{\imath_\nu\lro{\randvecc_\nu(w'), \randvecy_\nu}}=I(P_{\randvecx},P_{\randvecy\given\randvecx})>0$ for all $\nu\geq 1$, this random walk drifts to $+\infty$~\cite[Thm.~2.8.2]{gut09-a}.
 As a consequence, we conclude that
 \begin{IEEEeqnarray}{rCL}\label{eq:stopping_time_finite}
   \Pr\{\tau\sub{dec}<\infty\} &\geq & \Pr\{\tau_{w'}<\infty\}=1-\Pr\lefto\{\imath_\nu\lro{\randvecc^\nu(w'), \randvecy^\nu} <\gamma\sub{dec},\, \forall \nu \right\}=1.
 \end{IEEEeqnarray}
 Here, the first inequality follows from~\eqref{eq:tau_star} and the last equality follows because the random walk drifts to $+\infty$~\cite[Thm.~3.1.1]{gut09-a}.}

 We next prove that $\Ex{}{\ttx}$ can be upper-bounded as in~\eqref{eq:ell_ub}.
 Set $G_0=0$ and $G_\nu=\Ex{}{\ttx\given\tau\sub{dec}=\nu}$.
 One can show that for $\nu=1,\dots,\nmax-1$, the conditional expectation $G_\nu$ takes the form given in~\eqref{eq:h0-def}, whereas for $\nu\geq \nmax$
 \begin{IEEEeqnarray}{rCl}\label{eq:h0-beyond-lmax}
   G_{\nu} &=&\nmax(1-\markblue{\efb})^{\nmax-1} + \sum_{k=1}^{\nmax} k (1-\efb)^{k-1}\efb .
 \end{IEEEeqnarray}
 Note that this quantity does not depend on $\nu$.
 We next evaluate $\Ex{}{\ttx}$ as follows
 \markblue{
 \begin{IEEEeqnarray}{rCl}
   \Ex{}{\ttx} &=& \sum_{\nu=1}^{\infty} G_\nu \prob{\tau\sub{dec}=\nu}\\
   &=& \lim_{\ell\to\infty}\left(\sum_{\nu=1}^{\ell} G_\nu \bigl(\prob{\tau\sub{dec}>\nu-1}-\prob{\tau\sub{dec}>\nu}\bigr)\right)\\
   &=& \lim_{\ell\to\infty} \left( \sum_{\nu=0}^{\ell-1} (G_{\nu+1}-G_\nu) \prob{\tau\sub{dec}>\nu} - G_{\ell}\prob{\tau\sub{dec}>\ell} \right)\\
   &=& \sum_{\nu=0}^{\nmax-1} (G_{\nu+1}-G_\nu) \prob{\tau\sub{dec}>\nu}.
 \end{IEEEeqnarray}
 }
 In the last step we used~\eqref{eq:stopping_time_finite} and that $G_{\nu+1}=G_\nu$ for all $\nu\geq\nmax$ as a consequence of~\eqref{eq:h0-beyond-lmax}.
 Note now that $G_1>G_0$ by definition.
 Furthermore, standard algebraic manipulations reveal that, for $\nu=1,\dots,\nmax-1$,
 %
 \begin{IEEEeqnarray}{rCl}
    G_{\nu+1} - G_\nu 
 	&=&
 	\frac{\lro{1-\efa-\efb} \lro{1-\efb}^{\nu-1} \lro{1-\efa^{\nmax-\nu}}}{1-\efa} . \label{eq:Hj_diff}
 \end{IEEEeqnarray}
 This implies that $G_{\nu+1}-G_\nu \geq 0$ whenever $\efa+\efb\leq 1$.
 To obtain the desired result, we notice that
 \begin{equation}\label{eq:bound-on-taudec}
   \prob{\tau\sub{dec} > \nu} \leq \frac{1}{M}\sum_{w=1}^{M}\prob{\tau_w > \nu \given W=w}= \prob{\tau > \nu}
 \end{equation}
 where  $\tau$ is defined in~\eqref{eq:tau}.

 We now prove~\eqref{eq:eps_ub}.
 First note that, \markblue{since if threshold crossing occurs for more than one codeword, the one with largest index is chosen},
 \begin{IEEEeqnarray}{rCl}
   \epsilon &=& \frac{1}{M}\sum_{w=1}^M \prob{\widehat{W}\neq w\given W=w} \\
   &\leq& \prob{\widehat{W}\neq 1 \given W=1} \\
   &=& \sum_{\nu=1}^{\nmax} \prob{\trx=\nu,\widehat{W}\neq 1 \given W=1}.\label{eq:dec-err-prob-1}
 \end{IEEEeqnarray}
 Next, we decompose each term on the right-hand-side of~\eqref{eq:dec-err-prob-1}.
For $\nu=1$, the error probability coincides with the probability that an undetected error occurs in round $1$, i.e.,
\begin{IEEEeqnarray}{rCl}
     \prob{ \trx = 1, \widehat{W} \neq 1\given W=1}
&=&   \prob{\tau\sub{dec} = 1 , \widehat{W} \neq 1\given W=1}.\label{eq:splitting-error-prob_nu1}
\end{IEEEeqnarray}
For $\nu=2,\dots,\nmax-1$, we have
 %
 %
 \begin{IEEEeqnarray}{rCl}
  &&  \prob{ \trx = \nu, \widehat{W} \neq 1\given W=1} \notag \\
   &=&  \prob{\ttx = \nu-1, \tau\sub{dec} > \nu-1 , \widehat{W} \neq 1  \given W=1}   + \prob{\ttx \geq \nu, \tau\sub{dec} = \nu , \widehat{W} \neq 1 \given W=1}\IEEEeqnarraynumspace  \\
   &=& \prob{\ttx = \nu-1 \given  \tau\sub{dec} > \nu-1, W=1} \prob{\tau\sub{dec} > \nu-1 \given W=1}  \notag\\
   &&+ \prob{\ttx \geq \nu  \given  \tau\sub{dec} = \nu , \widehat{W} \neq 1, W=1}   \prob{\tau\sub{dec} = \nu, \widehat{W} \neq 1\given W=1}.\label{eq:splitting-error-prob}
 \end{IEEEeqnarray}
 The first term on the right-hand side of~\eqref{eq:splitting-error-prob} is the probability that an erasure is declared at step $\nu$ because of a $\nack\rightarrow \ack$ event at step $\nu-1$ and the second term on the right-hand side of~\eqref{eq:splitting-error-prob} corresponds to the probability of an undetected error.
 Observe now that
 \begin{IEEEeqnarray}{rCl}\label{eq:upper-bound-no-treshold-crossing}
   \prob{\tau\sub{dec} >\nu-1 \given W=1} &\leq& \prob{\tau  >\nu-1}.
 \end{IEEEeqnarray}
 Furthermore,
 %
 \begin{IEEEeqnarray}{rCl}
   \prob{\tau\sub{dec} = \nu, \widehat{W} \neq 1\given W=1}
   &=&
   \prob{\cup_{m=2}^M  \lrbo{\tau_1\geq \nu, \tau_m = \nu} \given W=1} \label{eq:proof_simp_a1}\\
   & \leq &
   \lro{M-1}  \prob{\tau_1\geq \nu, \tau_2 = \nu\given W=1} \label{eq:proof_simp_a2}\\
   & = &
   \lro{M-1}  \prob{\tau \geq \nu, \widetilde{\tau} = \nu} \label{eq:proof_simp_a3}
 \end{IEEEeqnarray}
 where $\widetilde{\tau}$ is defined in~\eqref{eq:taubar}.
 Finally,  we have that
 \begin{equation}\label{eq:bound-second-conditional-term}
   \prob{\ttx \geq \nu  \given  \tau\sub{dec} = \nu , \widehat{W} \neq 1, W=1} = \lro{1-\efb}^{\nu-1}
 \end{equation}
 and that
 \begin{equation}\label{eq:bound-first-conditional-term-nomax}
   \prob{\ttx = \nu-1 \given  \tau\sub{dec} > \nu-1, W=1} =\lro{1-\efb}^{\nu-2} \efb.
 \end{equation}
 For $\nu=\nmax$, the error probability is given by the sum of the terms in \eqref{eq:splitting-error-prob} computed for $\nu=\nmax$, and the additional term
  \begin{IEEEeqnarray}{rCl}\label{eq:bound-first-conditiona-term-max}
  \prob{\ttx \geq \nmax \given  \tau\sub{dec} > \nmax, W=1} \prob{\tau\sub{dec} > \nmax \given W=1}
  &\leq &
  \lro{1-\efb}^{\nmax-1}\prob{\tau >\nmax}.
  \end{IEEEeqnarray}
 This term describes the probability that no codeword causes a threshold crossing within $\nmax$ transmission rounds and no errors occurred on the feedback channel.
 We obtain the desired bound by substituting~\eqref{eq:upper-bound-no-treshold-crossing},~\eqref{eq:proof_simp_a3},~\eqref{eq:bound-second-conditional-term}, and~\eqref{eq:bound-first-conditional-term-nomax}
 into~\eqref{eq:splitting-error-prob_nu1} and~\eqref{eq:splitting-error-prob} and then~\eqref{eq:splitting-error-prob_nu1},~\eqref{eq:splitting-error-prob}, and~\eqref{eq:bound-first-conditiona-term-max} into~\eqref{eq:dec-err-prob-1}.
 %

 \section{Upper Bound on $\Ex{}{\trx}$ }\label{app:exp_tau_rx}
 Let $\trx$ be defined as in~\eqref{eq:tau_rx_def}.
 Furthermore, let $V_\nu=\Ex{}{\trx \given \tau\sub{dec}=\nu}$ for $\nu=1,2,\dots$ and $V_0 = 0$.
 The steps to bound $\Ex{}{\trx}$ are analogous to the ones used to bound $\Ex{}{\ttx}$ in Appendix~\ref{app:thm_ach_harq}.
 First, note that $V_1 = 1$.
 Next, we write
 \begin{IEEEeqnarray}{rCl}
 	\Ex{}{\trx} &=&
 	\sum_{\nu=0}^{\infty} \lro{ V_{\nu+1} - V_\nu} \prob{\tau\sub{dec} > \nu} \label{eq:ex_tilde_rx}
 \end{IEEEeqnarray}
 where
 \begin{IEEEeqnarray}{rCl}
   V_\nu &=&
  \nu \prob{\ttx \geq \nu \given \tau\sub{dec}=\nu} + \sum_{k=2}^{\nu} k \prob{\ttx = k-1 \given \tau\sub{dec} = \nu}   \\
     &=&
     \nu (1-\efb)^{\nu-1} +\sum_{k=2}^\nu k (1-\efb)^{k-2}\efb
 \end{IEEEeqnarray}
 for, $\nu=2,\dots,\nmax$, and
 \begin{IEEEeqnarray}{rCl}
   V_\nu &=& \nmax(1-\efb)^{\nmax-1} + \sum_{k=1}^{\nmax} k (1-\efb)^{k-2}\efb
 \end{IEEEeqnarray}
 for $\nu >\nmax$.
 Note that $V_1 > V_0$ and $V_{\nu+1}-V_\nu = 0$ for $\nu \geq \nmax$.
 Finally, for $\nu=1,\dots,\nmax-1$, we have
 \begin{IEEEeqnarray}{rCl}
 	V_{\nu+1} - V_\nu &=& (1-\efb)^{\nu-1} . \label{eq:Vj_diff}
 \end{IEEEeqnarray}
 Hence, we conclude that $V_{\nu+1}-V_\nu > 0$ and that
 \begin{IEEEeqnarray}{rCl}
 	\Ex{}{\trx} &\leq & 1 + \sum_{\nu=1}^{\nmax-1} (1-\efb)^{\nu-1} \prob{\tau > \nu} .
 \end{IEEEeqnarray}
 \section{Proof of \eqref{eq:simplification-bound-fading}} \label{app:HARQ_ach_mm}

By using Jensen's inequality in~\eqref{eq:generalized_info_dens}, we have that
 \begin{IEEEeqnarray}{rCl}
   \Ex{}{\jmath_\nu(\widetilde{\randvecx}^\nu, \randvecy^\nu)}
  & \leq & 0.
 \end{IEEEeqnarray}
 Since $\jmath_\nu(\widetilde{\randvecx}^\nu, \randvecy^\nu)$ is a sum of $\nu$ independent and identically distributed random variables, we conclude that each random variable has a negative mean.
 Such a property allows us to use Wald's identity~\cite[Cor. 9.4.4]{gallager13} and conclude that
 \begin{IEEEeqnarray}{rCl}
 \prob{ \widetilde{\tau} = \nu} & \leq & \prob{\jmath_\nu(\widetilde{\randvecx}^\nu, \randvecy^\nu) \geq \gamma\sub{dec}}\\
 &\leq &
 \exp\lro{-\beta^* \gamma\sub{dec}} \label{eq:wald-indentity}
 \end{IEEEeqnarray}
 Here, $\beta^*$ is the positive solution of
 \begin{IEEEeqnarray}{rCl}\label{eq:beta_optimization}
 \kappa\lro{\beta} = \log \Ex{}{\exp\lro{\beta \jmath_1(\widetilde{\randvecx}_1, \randvecy_1)}}
 &=&
 0.
 \end{IEEEeqnarray}
Substituting~\eqref{eq:generalized_info_dens_final} in~\eqref{eq:beta_optimization} we find that $\beta^*=1$.
Substituting this value in~\eqref{eq:wald-indentity}, we obtain the desired result.

  \bibliographystyle{IEEEtran}
  \bibliography{./giubib}

\begin{thebibliography}{10}
\providecommand{\url}[1]{#1}
\csname url@samestyle\endcsname
\providecommand{\newblock}{\relax}
\providecommand{\bibinfo}[2]{#2}
\providecommand{\BIBentrySTDinterwordspacing}{\spaceskip=0pt\relax}
\providecommand{\BIBentryALTinterwordstretchfactor}{4}
\providecommand{\BIBentryALTinterwordspacing}{\spaceskip=\fontdimen2\font plus
\BIBentryALTinterwordstretchfactor\fontdimen3\font minus
  \fontdimen4\font\relax}
\providecommand{\BIBforeignlanguage}[2]{{%
\expandafter\ifx\csname l@#1\endcsname\relax
\typeout{** WARNING: IEEEtran.bst: No hyphenation pattern has been}%
\typeout{** loaded for the language `#1'. Using the pattern for}%
\typeout{** the default language instead.}%
\else
\language=\csname l@#1\endcsname
\fi
#2}}
\providecommand{\BIBdecl}{\relax}
\BIBdecl

\bibitem{ostman19-08a}
J.~\"Ostman, R.~Devassy, G.~Durisi, and E.~G. Str\"om, ``On the nonasymptotic
  performance of variable-length codes with noisy stop feedback,'' in
  \emph{Proc. IEEE Inf. Theory Workshop (ITW)}, Visby, Sweden, Aug. 2019.

\bibitem{polyanskiy10-05a}
Y.~Polyanskiy, H.~V. Poor, and S.~Verd\'u, ``Channel coding rate in the finite
  blocklength regime,'' \emph{{IEEE} Trans. Inf. Theory}, vol.~56, no.~5, pp.
  2307--2359, May 2010.

\bibitem{yang14-07c}
W.~Yang, G.~Durisi, T.~Koch, and Y.~Polyanskiy, ``Quasi-static multiple-antenna
  fading channels at finite blocklength,'' \emph{{IEEE} Trans. Inf. Theory},
  vol.~60, no.~7, pp. 4232--4265, Jul. 2014.

\bibitem{durisi16-02a}
G.~Durisi, T.~Koch, J.~\"{O}stman, Y.~Polyanskiy, and W.~Yang, ``Short-packet
  communications over multiple-antenna {Rayleigh}-fading channels,''
  \emph{{IEEE} Trans. Commun.}, vol.~64, no.~2, pp. 618--629, Feb. 2016.

\bibitem{collins19-01}
A.~Collins and Y.~Polyanskiy, ``Coherent multiple-antenna block-fading channels
  at finite blocklength,'' \emph{{IEEE} Trans. Inf. Theory}, vol.~65, no.~1,
  pp. 380--405, Jan. 2019.

\bibitem{Ostman19-02}
J.~\"Ostman, G.~Durisi, E.~G. Str\"om, M.~C. Coskun, and G.~Liva, ``Short
  packets over block-memoryless fading channels: Pilot-assisted or noncoherent
  transmission?'' \emph{{IEEE} Trans. Commun.}, vol.~67, no.~2, pp. 1521--1536,
  Feb. 2019.

\bibitem{polyanskiy11-08a}
Y.~Polyanskiy, H.~V. Poor, and S.~Verd\'{u}, ``Feedback in the non-asymptotic
  regime,'' \emph{{IEEE} Trans. Inf. Theory}, vol.~57, no.~8, pp. 4903--4925,
  Aug. 2011.

\bibitem{bennis18-10}
M.~Bennis, M.~Debbah, and H.~V. Poor, ``Ultrareliable and low-latency wireless
  communication: Tail, risk, and scale,'' \emph{Proc. IEEE}, vol. 106, no.~10,
  pp. 1834--1853, Oct. 2018.

\bibitem{Shariatmadari18-06}
H.~Shariatmadari, S.~Iraji, R.~Jantti, P.~Popovski, Z.~Li, and M.~A. Uusitalo,
  ``Fifth-generation control channel design: Achieving ultrareliable
  low-latency communications,'' \emph{{IEEE} Veh. Technol. Mag.}, vol.~13,
  no.~2, pp. 84--93, Jun. 2018.

\bibitem{dahlman11-a}
E.~Dahlman, S.~Parkvall, and J.~Sk\"{o}ld, \emph{4G LTE/LTE-Advanced for Mobile
  Broadband}.\hskip 1em plus 0.5em minus 0.4em\relax Burlington, MA, U.S.A.:
  Elsevier, 2011.

\bibitem{shannon67-a}
C.~E. Shannon, R.~G. Gallager, and E.~R. Berlekamp, ``Lower bounds to error
  probability for coding on discrete memoryless channels---{Part {I}},''
  \emph{Inf. Contr.}, vol.~10, pp. 65--103, Feb. 1967.

\bibitem{gallager68a}
R.~G. Gallager, \emph{Information Theory and Reliable Communication}.\hskip 1em
  plus 0.5em minus 0.4em\relax New York, NY, U.S.A.: John Wiley \& Sons, 1968.

\bibitem{dobrushin62-a}
R.~L. Dobrushin, ``An asymptotic bound for the probability error of information
  transmission through a channel without memory using the feedback,''
  \emph{Problemy Kibernetiki}, vol.~8, pp. 161--168, 1961.

\bibitem{burnashev10-01}
M.~V. Burnashev and H.~Yamamoto, ``On the reliability function for a {BSC} with
  noisy feedback,'' \emph{Probl. Inf. Transm.}, vol.~46, no.~2, pp. 3--23, Jan.
  2010.

\bibitem{burnashev76-12a}
M.~V. Burnashev, ``{Data transmission over a discrete channel with feedback.
  Random transmission time},'' \emph{Probl. Inf. Transm.}, vol.~12, no.~4, pp.
  10--30, Dec. 1976.

\bibitem{Truong19}
L.~V. {Truong} and V.~Y.~F. {Tan}, ``Moderate deviation asymptotics for
  variable-length codes with feedback,'' \emph{{IEEE} Trans. Inf. Theory},
  vol.~65, no.~7, pp. 4364 -- 4386, Jul. 2019.

\bibitem{Draper08-04}
S.~C. Draper and A.~Sahai, ``Variable-length channel coding with noisy
  feedback,'' \emph{Eur. Trans. Telecommun.}, vol.~19, pp. 355--370, Apr. 2008.

\bibitem{Niesen09}
U.~Niesen and A.~Tchamkerten, ``Tracking stopping times through noisy
  observations,'' \emph{{IEEE} Trans. Inf. Theory}, vol.~55, no.~1, pp.
  422--432, Jan. 2009.

\bibitem{forney-jr68-03a}
G.~D. {Forney Jr}, ``Exponential error bounds for erasure, list, and decision
  feedback schemes,'' \emph{{IEEE} Trans. Inf. Theory}, vol.~14, no.~2, pp.
  206--220, Mar. 1968.

\bibitem{telatar92-05a}
I.~Telatar, ``Multi-access communication with decision feedback decoding,''
  Ph.D. dissertation, Massachusetts Institute of Technology, Cambridge, MA,
  USA, May 1992.

\bibitem{wu11-12}
P.~Wu and N.~Jindal, ``Coding versus {ARQ} in fading channels: How reliable
  should the {PHY} be?'' \emph{{IEEE} Trans. Commun.}, vol.~59, no.~12, pp.
  3363 -- 3374, Dec. 2011.

\bibitem{Avranas18}
A.~Avranas, M.~Kountouris, and P.~Ciblat, ``Energy-latency tradeoff in
  ultra-reliable low-latency communication with retransmissions,'' \emph{{IEEE}
  J. Sel. Areas Commun.}, vol.~11, no.~36, pp. 2475--2485, Nov. 2018.

\bibitem{Makki18-11}
B.~Makki, T.~Svensson, G.~Caire, and M.~Zorzi, ``Fast {HARQ} over finite
  blocklength codes: A technique for low-latency reliable communication,''
  \emph{{IEEE} Trans. Wireless Commun.}, vol.~18, no.~1, pp. 194--209, Jan.
  2018.

\bibitem{kim15-04}
S.~H. Kim, D.~K. Sung, and T.~Le-Ngoc, ``Variable-length feedback codes under a
  strict delay constraint,'' \emph{{IEEE} Commun. Lett.}, vol.~19, no.~4, pp.
  513--516, Apr. 2015.

\bibitem{williamson15-07a}
A.~R. Williamson, T.-Y. Chen, and R.~D. Wesel, ``Variable-length convolutional
  coding for short blocklengths with decision feedback,'' \emph{{IEEE} Trans.
  Commun.}, vol.~63, no.~7, pp. 2389--2403, Jul. 2015.

\bibitem{ostman18-12}
J.~\"{O}stman, R.~Devassy, G.~C. Ferrante, and G.~Durisi, ``Low-latency
  short-packet transmissions: Fixed length or {HARQ}?'' in \emph{Proc. IEEE
  Global Telecommun. Conf. (GLOBECOM)}, Abu Dhabi, UAE, Dec. 2018.

\bibitem{rahul19-04}
R.~Devassy, G.~Durisi, G.~C. Ferrante, O.~Simeone, and E.~Uysal, ``Reliable
  transmission of short packets through queues and noisy channels under latency
  and peak-age violation guarantees,'' \emph{{IEEE} J. Sel. Areas Commun.},
  vol.~4, no.~37, pp. 721--734, Apr. 2019.

\bibitem{Trillinsgaard18-12}
K.~F. Trillingsgaard, W.~Yang, G.~Durisi, and P.~Popovski, ``Common-message
  broadcast channels with feedback in the nonasymptotic regime: Stop
  feedback,'' \emph{{IEEE} Trans. Inf. Theory}, vol.~64, no.~12, pp.
  7686--7718, Dec. 2018.

\bibitem{Malkamaki00-09}
E.~Malkam\"{a}ki and H.~Leib, ``Performance of truncated type-{II} hybrid {ARQ}
  schemes with noisy feedback over block fading channels,'' \emph{{IEEE} Trans.
  Commun.}, vol.~48, no.~9, pp. 1477--1487, Sep. 2000.

\bibitem{neyman33-01a}
J.~Neyman and E.~S. Pearson, ``On the problem of the most efficient tests of
  statistical hypotheses,'' \emph{Phil. Trans. Roy. Soc. A}, vol. 231, pp.
  289--337, Jan. 1933.

\bibitem{Popovski18}
P.~{Popovski}, J.~J. {Nielsen}, C.~{Stefanovi\'{c}}, E.~d.~{Carvalho}, E.~G.
  {Str\"{o}m}, K.~F. {Trillingsgaard}, A.-S. {Bana}, D.~M. {Kim}, R.~{Kotaba},
  J.~{Park}, and R.~B. {S{\o}rensen}, ``Wireless access for ultra-reliable
  low-latency communication: Principles and building blocks,'' \emph{IEEE
  Network Magazine}, vol.~32, no.~2, pp. 16--23, Mar. 2018.

\bibitem{font-segura18-03}
J.~Font-Segura, G.~Vazquez-Vilar, A.~Martinez, A.~Guill\'{e}n~i F\`{a}bregas,
  and A.~Lancho, ``Saddlepoint approximations of lower and upper bounds to the
  error probability in channel coding,'' in \emph{Proc. Conf. Inf. Sci. Sys.
  (CISS)}, Princeton, NJ, USA, Mar. 2018.

\bibitem{martinez11-02a}
A.~Martinez and A.~{Guill{\'e}n i F{\`a}bregas}, ``Saddlepoint approximation of
  random--coding bounds,'' in \emph{Proc. Inf. Theory Applicat. Workshop
  (ITA)}, San Diego, CA, U.S.A., Feb. 2011.

\bibitem{wu10-04}
P.~Wu and N.~Jindal, ``Performance of hybrid-{ARQ} in block-fading channels: A
  fixed outage probability analysis,'' \emph{{IEEE} Trans. Commun.}, vol.~58,
  no.~4, pp. 1129--1141, Apr. 2010.

\bibitem{gut09-a}
A.~Gut, \emph{Stopped random walks: limit theorems and applications},
  2nd~ed.\hskip 1em plus 0.5em minus 0.4em\relax New York, NY, USA: Springer,
  2009.

\bibitem{gallager13}
R.~G. Gallager, \emph{Stochastic Processes: Theory for Applications}.\hskip 1em
  plus 0.5em minus 0.4em\relax Cambridge, U.K.: Cambridge Univ. Press, 2013.

\end{thebibliography}
 \end{document}